\documentclass[a4paper,11pt]{article}
\usepackage{jheppub} 
\usepackage{multirow}
\usepackage{rotating}
\usepackage{lineno}

\title{\boldmath New Insights on Gamma-Ray Burst Radiation Mechanisms from Multiwavelength Observations}

\author[1,2,3]{Yu-Hua Yao}
\author[2]{Fang-Sheng Min}
\author[4,2]{Shi Chen}
\author[2,5,6]{Yi-Qing Guo}

\affiliation[1]{College of Physics, Chongqing University,
No.55 Daxuecheng South Road, High-tech District, Chongqing, 401331, China}
\affiliation[2]{Key Laboratory of Particle Astrophysics, Institute of High Energy Physics, Chinese Academy of Sciences, Beijing 100049, China}
\affiliation[3]{Wisconsin IceCube Particle Astrophysics Center, University of Wisconsin–Madison, Madison, WI 53703, USA}

\affiliation[4]{School of Physics and Astronomy, Yunnan University, Yunnan 650091, China}
\affiliation[5]{
College of Physics, University of Chinese Academy of Sciences, 100049, Beijing, China
}
\affiliation[6]{Tianfu Cosmic Ray Research Center, Chengdu 610213, Sichuan Province, China}

\emailAdd{yyao255@wisc.edu, guoyq@ihep.ac.cn}

\abstract{The study of high-energy gamma-ray emission from gamma-ray bursts (GRBs) involves complex synchrotron radiation and synchrotron self-Compton scattering (SSC) mechanisms with multiple parameters exhibiting a wide distribution. Recent advancements in GRB research, particularly the observation of very high energy (VHE, $\rm >100~GeV$) radiation, have ushered in a new era of multiwavelength exploration, offering fresh perspectives and limitations for understanding GRB radiation mechanisms. This study aimed to leverage VHE observations to refine constraints on synchrotron + SSC radiation from electrons accelerated by forward shocks. By analyzing two external environments - the uniform interstellar medium and stratified stellar wind medium, we conducted spectral and variability fitting for five specific bursts (GRB~180720B, GRB~190114C, GRB~190829A, GRB~201216C, and GRB~221009A) to identify the optimal parameters characterizing these events. A comparative analysis of model parameter distributions with and without VHE radiation observations reveals that the magnetic energy equipartition factor $\epsilon_B$ is more concentrated with VHE emissions. This suggests that VHE emissions may offer greater constraints on this microphysical parameter. Additionally, we found that the energy budget between VHE and keV-MeV $\gamma$-ray emissions under the SSC radiation exhibits an almost linear relationship, which may serve as a tool to differentiate radiation mechanisms. We anticipate future statistical analyses of additional VHE bursts to validate our findings.}

\begin{document}
\maketitle

\flushbottom

\section{Introduction} \label{sec:intro}

Gamma-ray bursts (GRBs) stand out as the most energetic and intense celestial phenomena, manifesting unpredictably across the universe. Since their initial discovery in the 1960s, satellites and ground-based experiments have documented a multitude of GRB occurrences, spanning a broad spectrum from prompt emissions to afterglow radiation, encompassing various wavelengths including radio, optical, X-ray, and extending up to GeV gamma-ray energies \citep{2016ApJ...829....7L,2019ApJ...878...52A,2020ApJ...893...46V}. The theoretical prediction \citep{1998ApJ...494L.167P,2000ApJ...537..785D} of TeV radiation and the extensive experimental searches conducted over numerous decades culminated in the breakthrough achievement of MAGIC in opening the very high energy (VHE, $\rm >100~GeV$) window \citep{2019Natur.575..455M}, marking a significant milestone in the formal inclusion of GRB research within the complete electromagnetic spectrum.

VHE radiation observations have injected new vitality into the study of GRBs ( see reviews \cite{2022Galax..10...66M,2022Galax..10...67B} and referers therein). Synchrotron radiation is widely regarded as the most natural explanation for GRB sub-MeV emission \citep{1999ApJ...523..177W,2001ApJ...548..787S}. The observation of VHE radiation in GRB 190114C confirmed the presence of components beyond synchrotron radiation in the afterglow phase for the first time at energies near TeV \citep{2019Natur.575..459M}. This finding provides further support to earlier observations by Fermi-LAT of an additional component \citep{2014Sci...343...42A}, and also bolsters the possibility that inverse Compton emission is commonly produced in GRBs. Inverse Compton radiation from relativistic electrons is commonly used to explain such high-energy photons \citep{2001ApJ...548..787S}, while hadronic models have also been proposed for a comprehensive explanation \citep{2022ApJ...929...70S,2023A&A...670L..12D}, along with exotic origins \citep{2023PhRvL.131y1001G}. Additionally, observations of VHE radiation have imposed new constraints on the study of galactic and extragalactic magnetic fields \citep{2023ApJ...955L..10H,2024MNRAS.527L..95D}, Lorentz invariance \citep{2023PhRvD.107h3001Z,2023ApJ...942L..21F}, and other related aspects.

Many aspects of GRBs, such as the particle acceleration mechanism, the role of inverse Compton emission, and the properties of the environment, remain poorly constrained. Model studies have revealed that constraints on model microphysics parameters derived from previous GRB data are not universal and are distributed over a wide range \citep{2014ApJ...785...29S}. The inclusion of the VHE emission might enrich the study of GRBs. To date, since limited VHE GRB observed, there are only a limited number of articles employed a consistent model for the analysis of these VHE GRBs \citep{2019ApJ...884..117W,2022Galax..10...66M,2023MNRAS.523..149G,2024arXiv240313902K}, and discrepancies and variations among radiation models pose challenges when attempting comparisons between bursts. Therefore, this paper aims to use the forward shock model to describe the existing VHE-emission bursts.

In this paper, we conduct a multi-wavelength analysis using the synchrotron + SSC model to study the light curves and spectral energy distributions (SEDs) of VHE GRB afterglows. Our aim is to investigate their common properties and examine the physical conditions in the emitting region of VHE GRB afterglows. This paper is organized as follows: Section 2 presents the afterglow model used for modeling, followed by the results in Section 3. Finally, Section 4 provides a summary.

\section{Models and Methods} 

Employing the dynamic evolution model for GRB afterglows presented in \citep{2000ApJ...543...90H}, the SEDs and light curves of the GRB afterglows are obtained by considering the afterglow emission is produced by electrons accelerated in the forward shocks. Since many works have explained the broadband SEDs and light curves of GRB afterglows, here we give a simple description about essential processes, consisting of dynamical evolution, shock-accelerated electron distribution, radiation process first, for more details about the model description see \cite{2001ApJ...548..787S, 2008MNRAS.384.1483F,2013ApJ...773L..20L}.

\subsection{Dynamical Evolution}
An impulsive relativistic outflow, characterized by an initial kinetic energy $E_{k}=E_0$ and an initial Lorentz factor $\Gamma_{0}$, propagates into an external medium with a density profile $n=n_0~(R/R_0)^{-k}$. For $k=0$, the medium exhibits a constant density $n=n_{0}$ in the interstellar medium (ISM) scenario, while $k=2$ corresponds to a density profile $n=n_0 (R/R_0)^{-2}$ in the stellar-wind case. The overall dynamic evolution of the radius R, Lorentz factor $\Gamma$, and energy $E_k$ can be described by the following equations \citep{2000ApJ...543...90H}

\begin{equation}
\frac{d R}{d T}=\frac{\sqrt{1-\frac{1}{\Gamma_s^2} c \Gamma}}{\Gamma-\sqrt{\Gamma^2-1}}
\end{equation}

\begin{equation}
\frac{d \Gamma}{d R}=\frac{-4 \pi n m_p R^2\left(\Gamma^2-1\right)}{\frac{E_0}{\Gamma_0 c^2}+(\epsilon+2 \Gamma(1-\epsilon))\left(\frac{4 \pi R^3 n m_p}{3-k}+\frac{E_0}{4 \Gamma_0 c^2}\right)}
\end{equation}

\begin{equation}
\frac{d E_k}{d R}=-\epsilon \Gamma(\Gamma-1) 4 \pi R^2 m_p n c^2
\end{equation}
Here, $m_p$ is the mass of proton, $c$ is the speed of light. $\epsilon$ is the radiative efficiency defined as the fraction of the shock generated thermal energy that is
radiated, $\epsilon=1$ corresponds to highly radiative case, and $\epsilon=0$ corresponds to adiabatic expansion. $\Gamma_s$ denotes the Lorentz factor of the shock related to the bulk Lorentz factor with the adiabatic index $\hat{\gamma}=(4\Gamma+1)/3\Gamma$, as derived by \cite{1976PhFl...19.1130B}
\begin{equation}
     \Gamma_s=\sqrt{\frac{(\Gamma+1)(\hat{\gamma}(\Gamma-1)+1)^2}{\hat{\gamma}(2-\hat{\gamma})(\Gamma-1)+2}}
\end{equation}

\subsection{Shock-accelerated Electrons}
Relativistic shocks serve as sites for particle acceleration, magnetic field amplification, and photon radiation \citep{zhang2018physics}. In the context of these shocks, the fractions of shock internal energy allocated to electrons and magnetic fields are represented by the constants $\epsilon_e$ and $\epsilon_B$ respectively. The distribution of accelerated electrons is commonly described by a single power-law function. In the slowing cooling case the distribution is given by \citep{1998ApJ...497L..17S}
\begin{equation}
\frac{d N_e}{d \gamma_e} \propto \begin{cases}\left(\frac{\gamma_e}{\gamma_c}\right)^{-p}, & \gamma_m \leqslant \gamma_e \leqslant \gamma_c \\ \left(\frac{\gamma_e}{\gamma_c}\right)^{-p-1}, & \gamma_c \leqslant \gamma_e \leqslant \gamma_{\max }\end{cases}
\end{equation}
and in the fast cooling case it goes with
\begin{equation}
\frac{d N_e}{d \gamma_e} \propto \begin{cases}\left(\frac{\gamma_e}{\gamma_m}\right)^{-2}, & \gamma_c \leqslant \gamma_e \leqslant \gamma_m \\ \left(\frac{\gamma_e}{\gamma_m}\right)^{-p-1}, & \gamma_m \leqslant \gamma_e \leqslant \gamma_{M}\end{cases}
\end{equation}
$\gamma_{e}, \gamma_m$ and $\gamma_M$ are the electron Lorentz factor, minimum and maximum Lorentz factors \citep{1976PhFl...19.1130B,1998ApJ...497L..17S}.
\begin{equation}\label{eq_gamma_m}
\gamma_{m} = \frac{p-2}{p-1} \epsilon_e (\Gamma-1) \frac{m_p}{m_e},
\end{equation}

\begin{equation}\label{eq_gamma_M}
\gamma_M = \sqrt{\frac{6\pi e}{\sigma_TB'(1+Y)}}
\end{equation}
$\sigma_T$ is the Thomson scattering cross-section, $m_e$ and $m_p$ are the mass of electron and proton, respectively. The magnetic field intensity \citep{2001ApJ...548..787S,2005MNRAS.363.1409P} is parameterized as
\begin{equation}
 B'=\sqrt{8\pi \frac{\epsilon_{B}(\hat{\gamma}\Gamma_s+1)(\Gamma_s-1)}{\hat{\gamma}-1}nm_pc^2}.
\end{equation}
Primed values are calculated in the co-moving frame of the jet. Considering the dynamical time scale, the cooling Lorentz factor of electrons $\gamma_c$ can be described as:
\begin{equation}\label{eq_gamma_c}
\gamma_c = \frac{6\pi m_e c}{\sigma_TB'^{2}(1+Y)t}
\end{equation}

Here, the parameter Y evaluates the effect of SSC cooling on the synchrotron spectrum \citep{2001ApJ...548..787S}. Considering the simplest cast with first-oder SSC only following \citep{2001ApJ...548..787S}, one has
\begin{equation}
Y \equiv \frac{L_{\mathrm{SSC}}}{L_{\mathrm{syn}}}=\frac{U_{\mathrm{syn}}}{U_B}=\frac{\eta_e U_e /(1+Y)}{U_B}=\frac{\eta_e \epsilon_e}{\epsilon_B(1+Y)},
\end{equation}
where $\eta_e$ is the electron radiation efficiency with
\begin{equation}
\eta_e= \begin{cases}1, & \text { fast cooling } \\ \left(\frac{\gamma_c}{\gamma_m}\right)^{2-p}, & \text { slow cooling }.\end{cases}
\end{equation}
Thus one gets
\begin{equation}
Y=\frac{-1+\sqrt{1+4 \eta_e \epsilon_e / \epsilon_B}}{2}
\end{equation}
Klein–Nishina suppression and Doppler effect have been taken into account, while the equal-arrival-time effect is ignored for simplicity.

\subsection{Synchrotron and SSC Radiation}
The accelerated electrons, driven by shock waves, lose energy through radiation within the magnetic field of the shock-heated region. The power of synchrotron radiation can be calculated using the traditional formula proposed by \citep{1986A&A...164L..16C}:
\begin{equation}
P_{\text{syn}}'(\nu') = \frac{\sqrt{3}e^3B'}{m_ec^2}\int d\gamma_e'\frac{dN_e'}{d\gamma_e'}F\left(\frac{\nu'}{\nu_{\text{ch}}'}\right),
\end{equation}
where $\nu_{\text{ch}}' = \frac{3\gamma'^2eB'\sin\alpha}{4\pi m_ec}$ represents the characteristic frequency of photons, and $F(x)$ denotes the synchrotron function containing the modified Bessel function.

When relativistic electrons interact with a photon field, they undergo scattering events, transferring their kinetic energy to the photons. The cross section for this scattering process can be expressed \citep{Rybicki1979RadiativePI} as
\begin{equation}
\begin{split}
  \sigma=\frac{3 \sigma_T}{4} \left\{ \frac{1+x}{x^3}\left[\frac{2 x(1+x)}{1+2 x}-\ln (1+2 x)\right] \right. \\
        +\left. \frac{1}{2 x} \ln (1+2 x)-\frac{1+3 x}{(1+2 x)^2}\right\}
\end{split}
\end{equation}
$x \equiv \frac{h v}{m_e c^2}$, and nne can see that $\sigma\sim\sigma_{T}$ in the Thomson regime ($x<<1$), and $\sigma$ is greatly suppressed ($\propto \sigma_T/x$) in the Klein–Nishina regime ($x>>1$).

Similar to the synchrotron process, the total spectral energy distribution of the SSC radiation can be described  \citep{2008MNRAS.384.1483F} as
\begin{equation} P_{\text{ssc}}'(\nu_{\text{ssc}}') = \int\int \frac{dN_e'}{d\gamma_e'}h\nu_{\text{ssc}}' \frac{dN_{\gamma}'}{dt'd\nu_{\text{ssc}}'}d\gamma_e' \end{equation}
with
\begin{equation} \frac{dN_{\gamma}'}{dt'd\nu_{\text{ssc}}'} = \frac{3\sigma_Tc}{4\gamma_e'^2}\frac{n_{\nu'}d\nu'}{\nu'} \end{equation}
Here, $F(q,g)=2qlnq+(1+2q)(1-q)+\frac{(4qg)^2}{2(1+4gq)}(1-q)$.

The intrinsic spectral flux in the observer frame can be derived from the radiation power using the following equation:
\begin{equation} F(\nu, t) = \frac{(1+z)P'(\nu'(\nu))\Gamma}{4\pi D_L^{2}}. \end{equation}
In this equation, $ D_L$ represents the luminosity distance between the source and the observer. The cosmological constant $ H_0 = 71~km~s^{-1}~Mpc^{-1}$, $ \Omega_M = 0.3$, and  $ \Omega_\lambda = 0.7$.

High-energy $\gamma$-rays, due to the process $\gamma\gamma\rightarrow e^{+}e^{-}$, may experience internal absorption by ambient photons within the source. This absorption is corrected by a flux correction factor of $\frac{(1-e^{-\tau_i})}{\tau_i}$. Additionally, external absorption by the extragalactic background light (EBL) occurs, resulting in an attenuation factor of $e^{-\tau_{EBL}}$. In this study, we utilize the EBL model developed by \citep{2021MNRAS.507.5144S}, which is favored
by LHAASO for the observation of GRB221009A.

In short, the observed frequency $F(\nu, t)$ for an individual GRB is determined by a specific set of model parameters, including $ E_k, \Gamma_0, \epsilon_e, \epsilon_B, p, z, n_0$. In this research, we employed the numerical code for afterglow modeling established by \cite{2013ApJ...773L..20L}.

\begin{figure*}[!htp]
\centering
\includegraphics[width=0.42\linewidth]{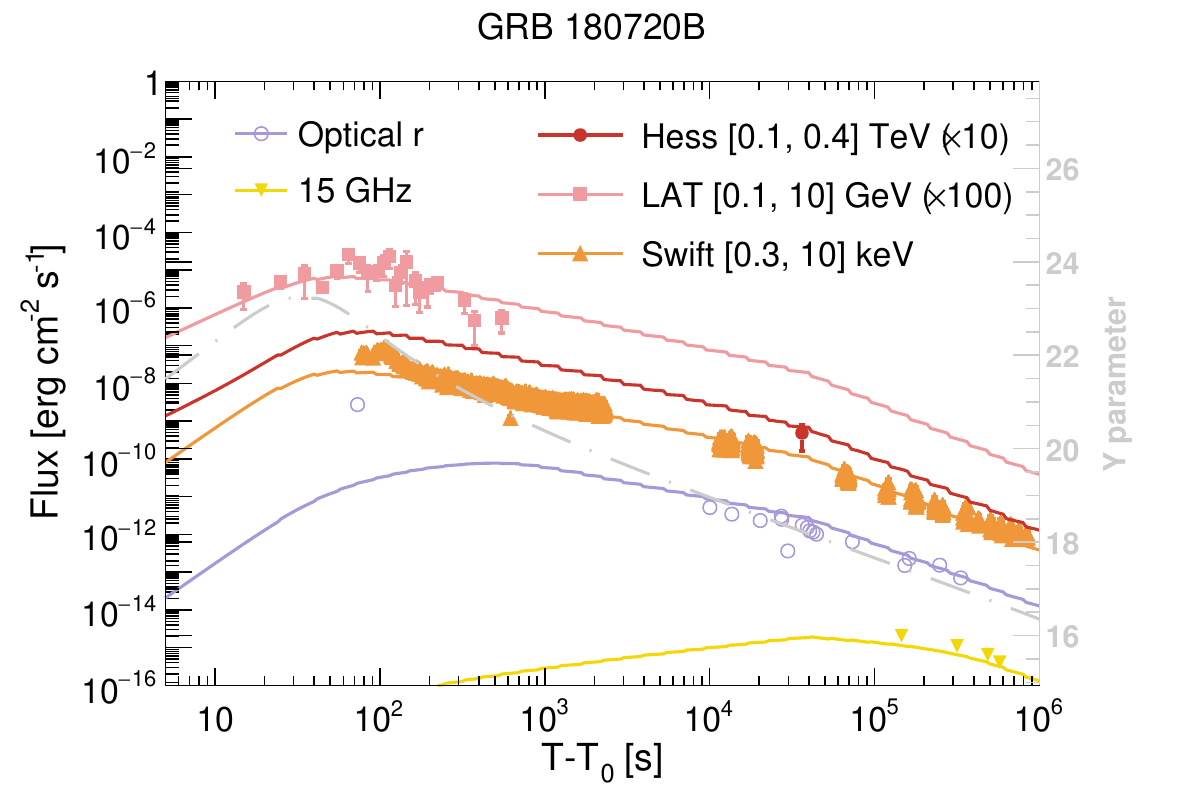}
\includegraphics[width=0.42\linewidth]{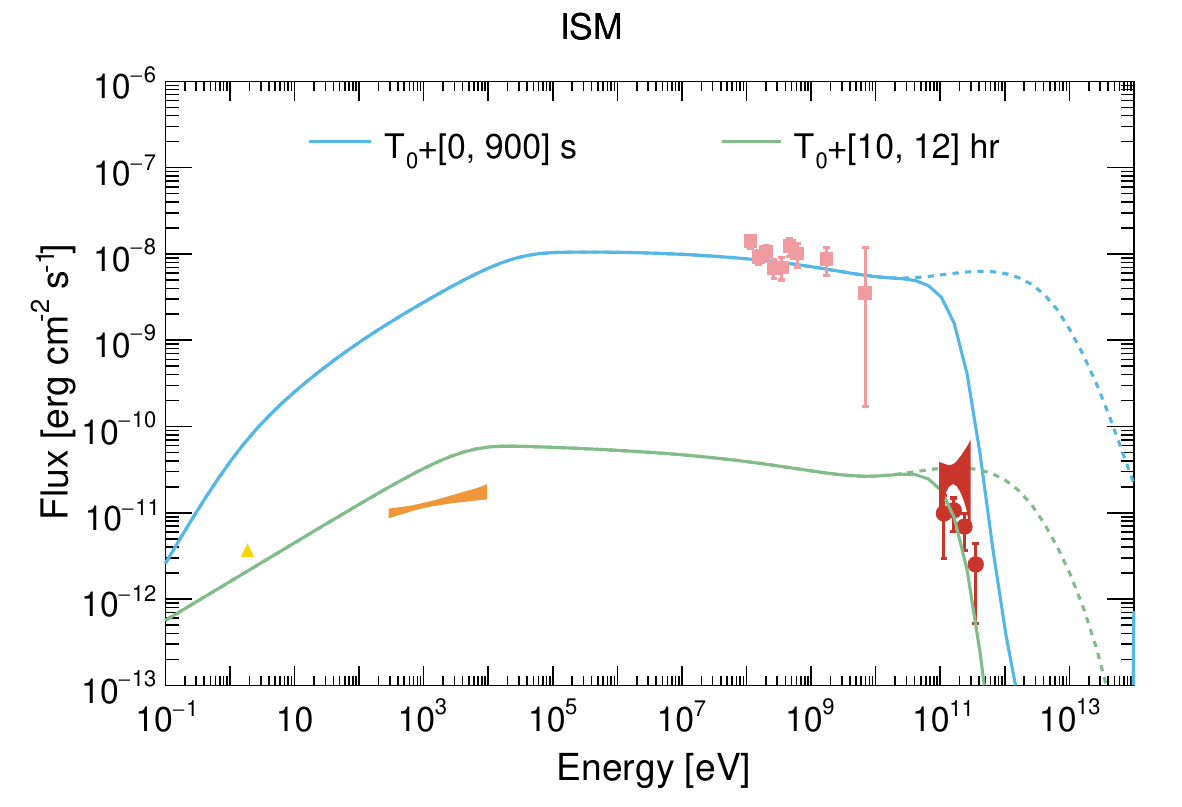}
\includegraphics[width=0.42\linewidth]{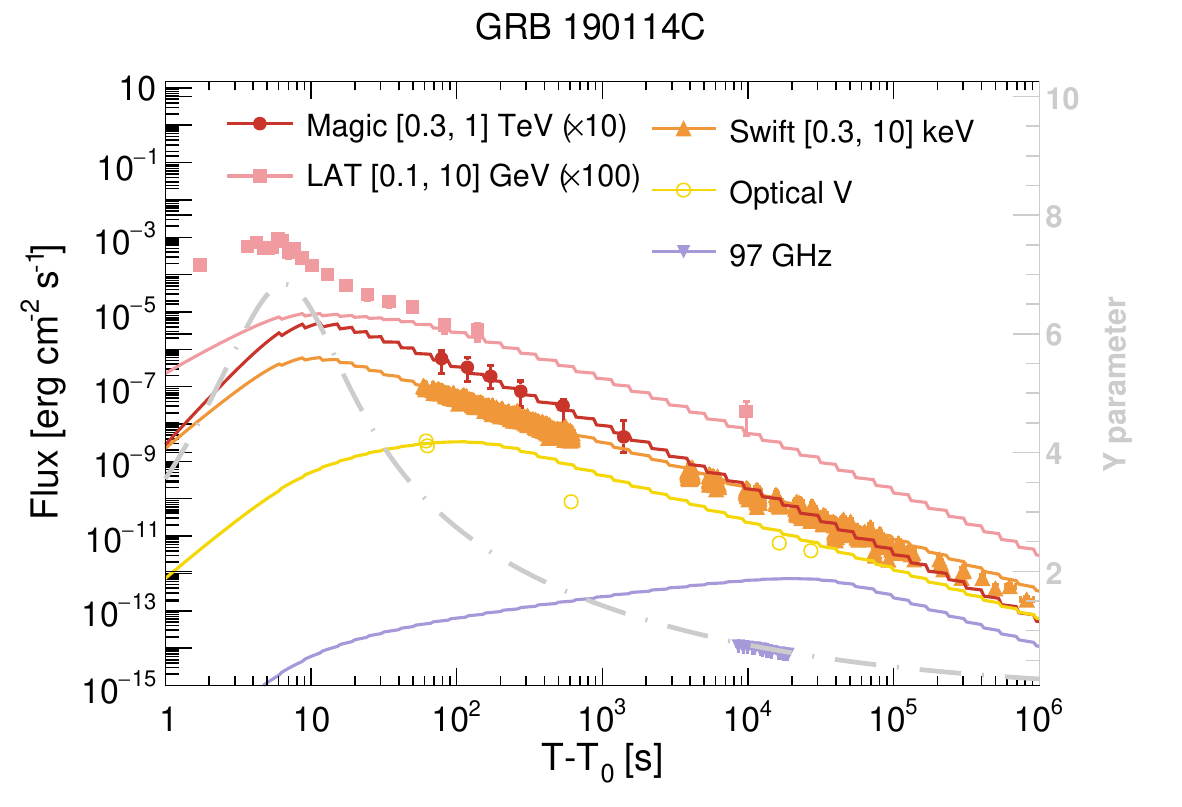}
\includegraphics[width=0.42\linewidth]{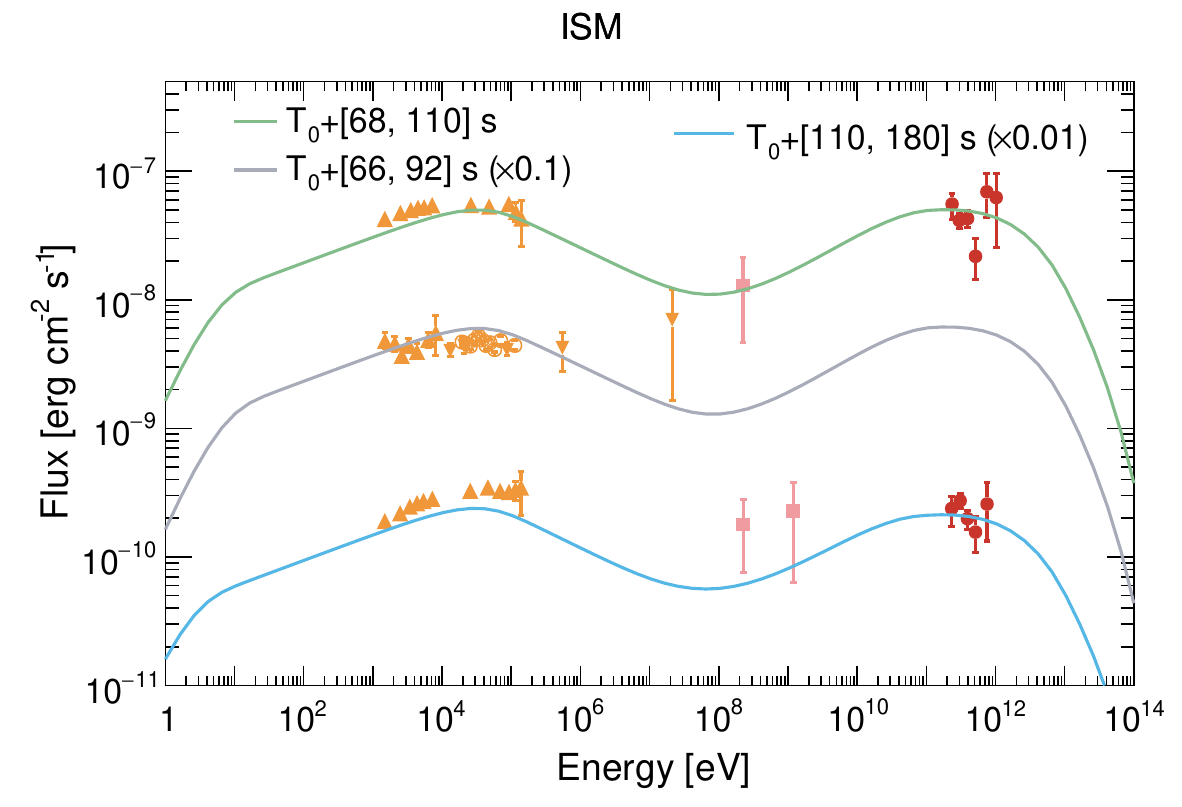}
\includegraphics[width=0.42\linewidth]{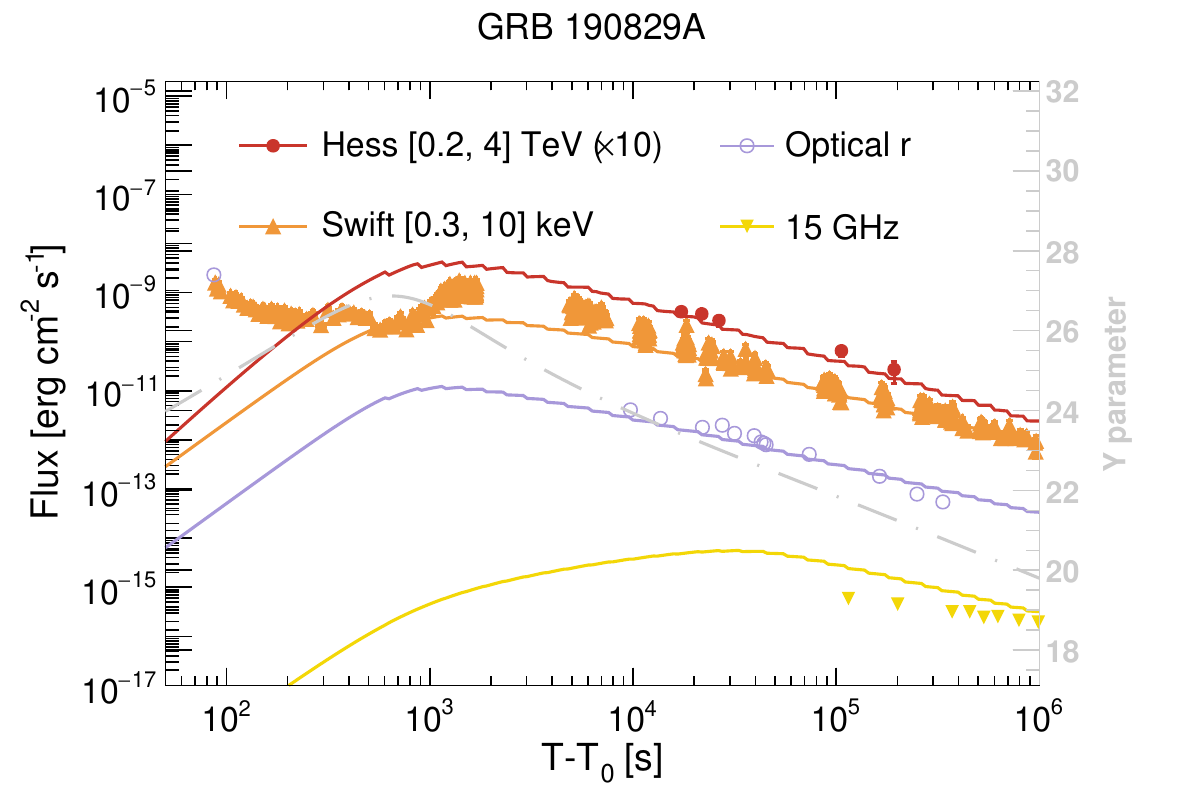}
\includegraphics[width=0.42\linewidth]{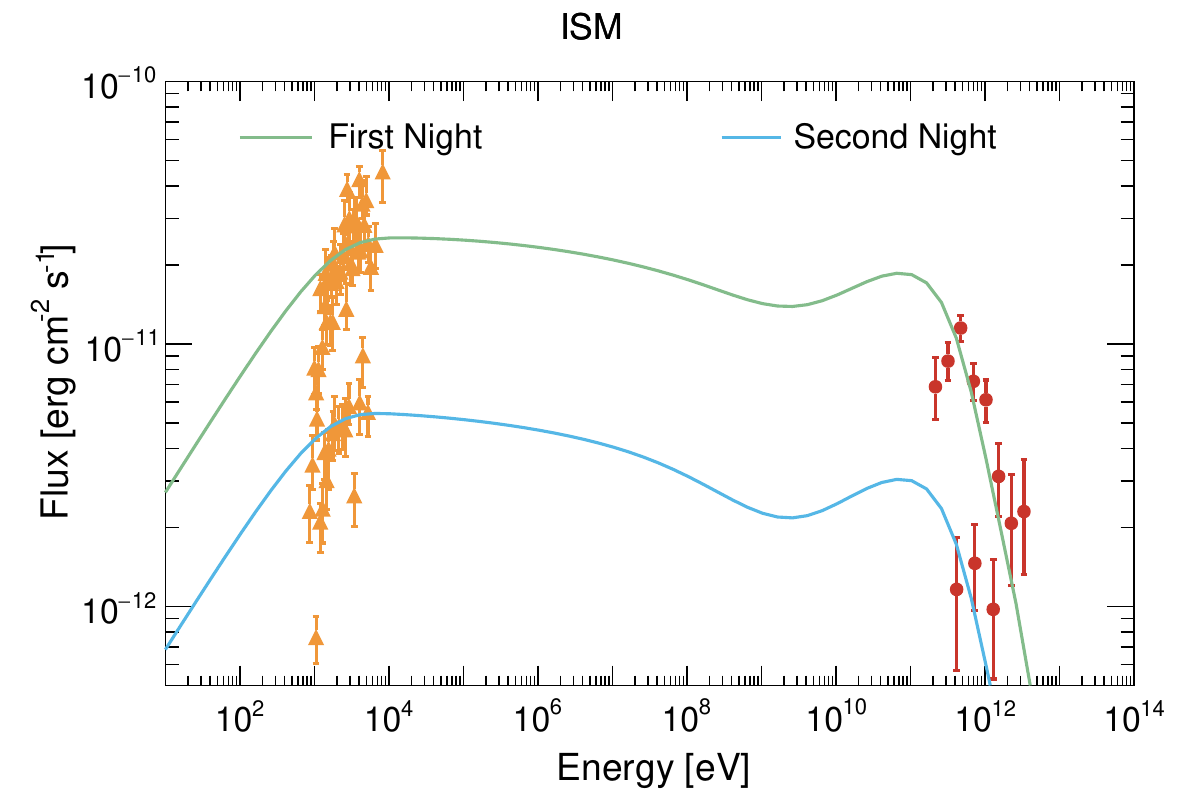}
\includegraphics[width=0.42\linewidth]{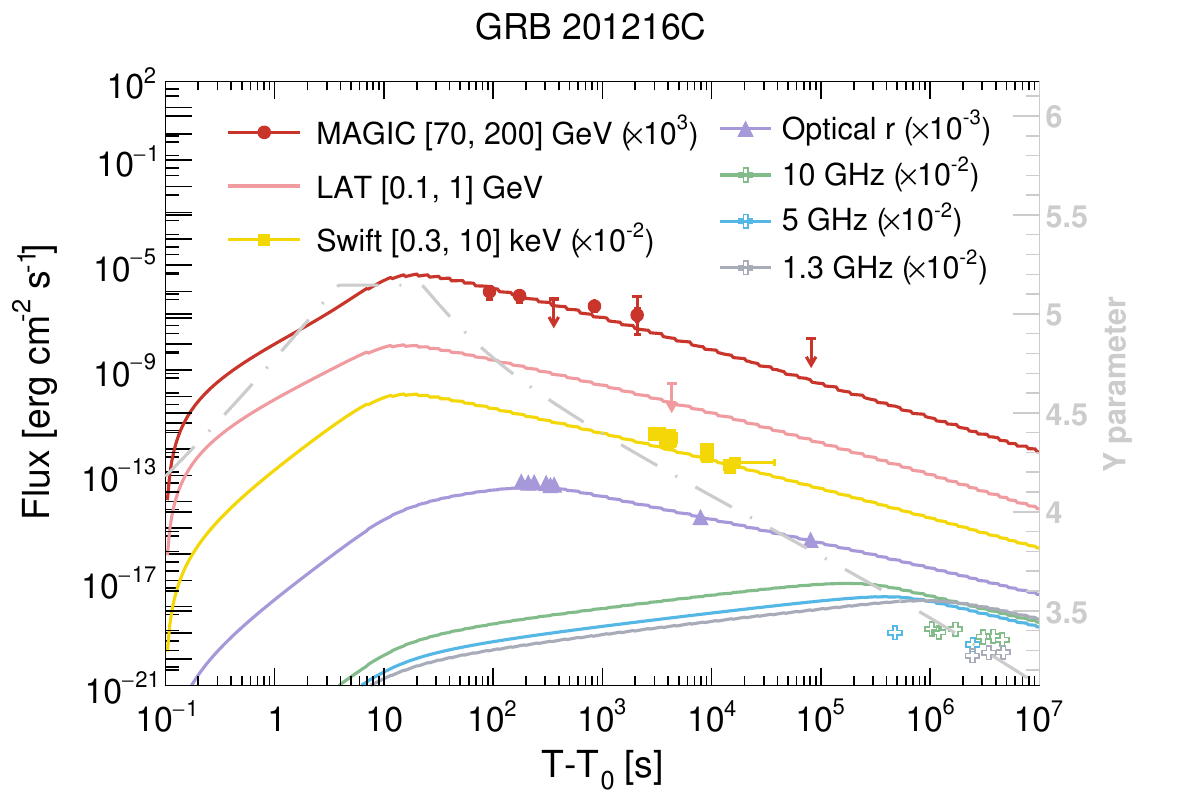}
\includegraphics[width=0.42\linewidth]{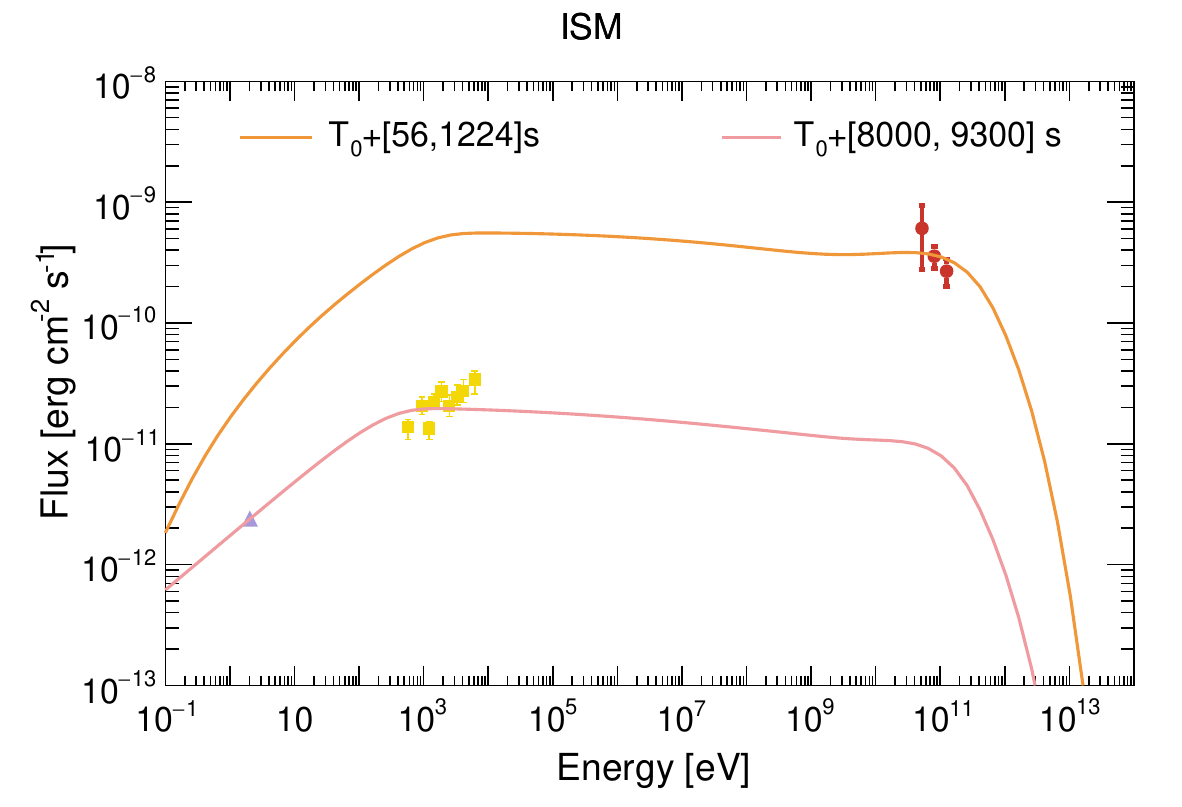}
\includegraphics[width=0.42\linewidth]{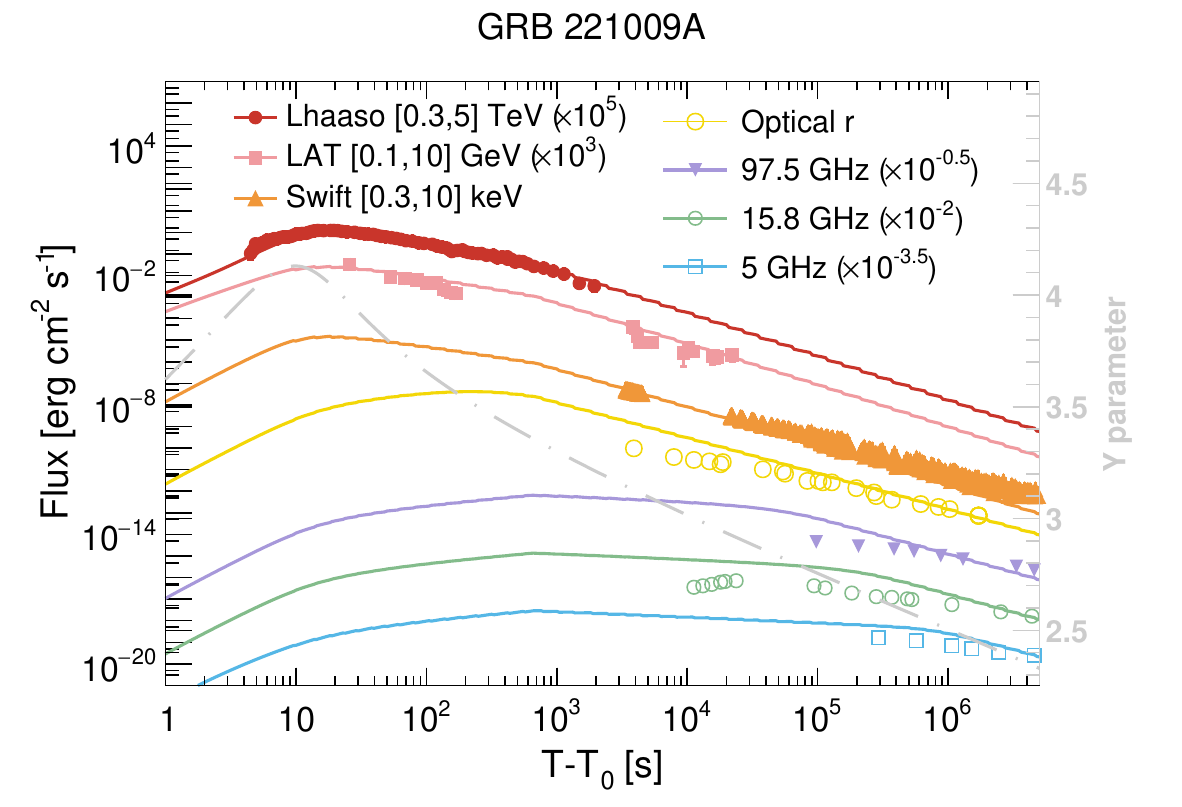}
\includegraphics[width=0.42\linewidth]{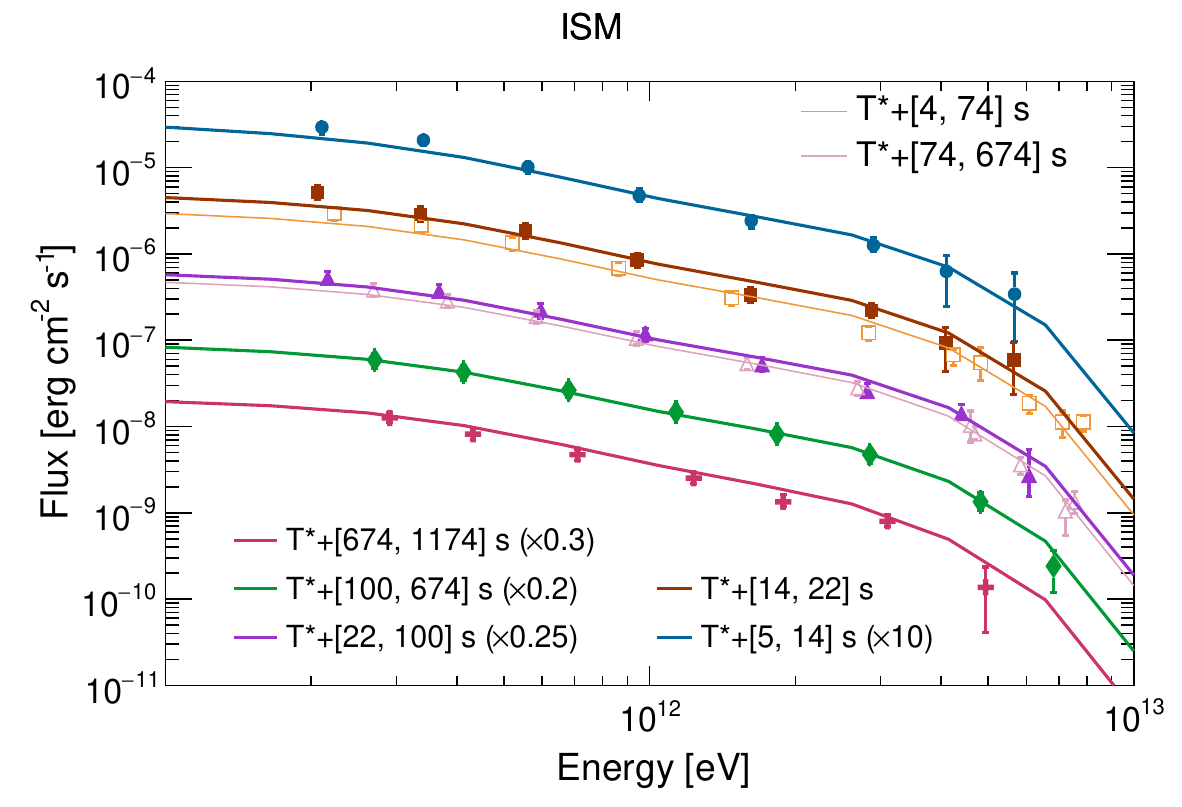}
\caption{{\scriptsize{Light curves (left) and SEDs (right) of SSC emission expected from a decelerating outflow in a homogeneous medium. Lines in different colors are calculated with models, markers are observational data from different experiments. In the right panels, the dashed line in the highest energy range represents the flux after correcting EBL absorption. The observational data for GRB 180720B are obtained from \cite{2019Natur.575..464A,2019ApJ...885...29F,2019ApJ...885...29F,2024NatAs...8..134A}, for GRB 190114C are sourced from \cite{2019Natur.575..455M,2019ApJ...878L..26L,2019ApJ...883..162F}, for GRB 201216C are from \cite{2024MNRAS.527.5856A}, for GRB 190829A are obtained from \cite{2021Sci...372.1081H,2023ApJ...947...84H}, and for GRB 221009A are sourced from \cite{2023Sci...380.1390L,2023SciA....9J2778C,2024ApJ...962..115R}. The gray dashed lines in the left panel represent the time evolution of the Y parameter, their coordinates are the right y-axis of the graph.
}}}
\label{fig:const}
\end{figure*}

\begin{figure*}[!htp]
\centering
\includegraphics[width=0.42\linewidth]{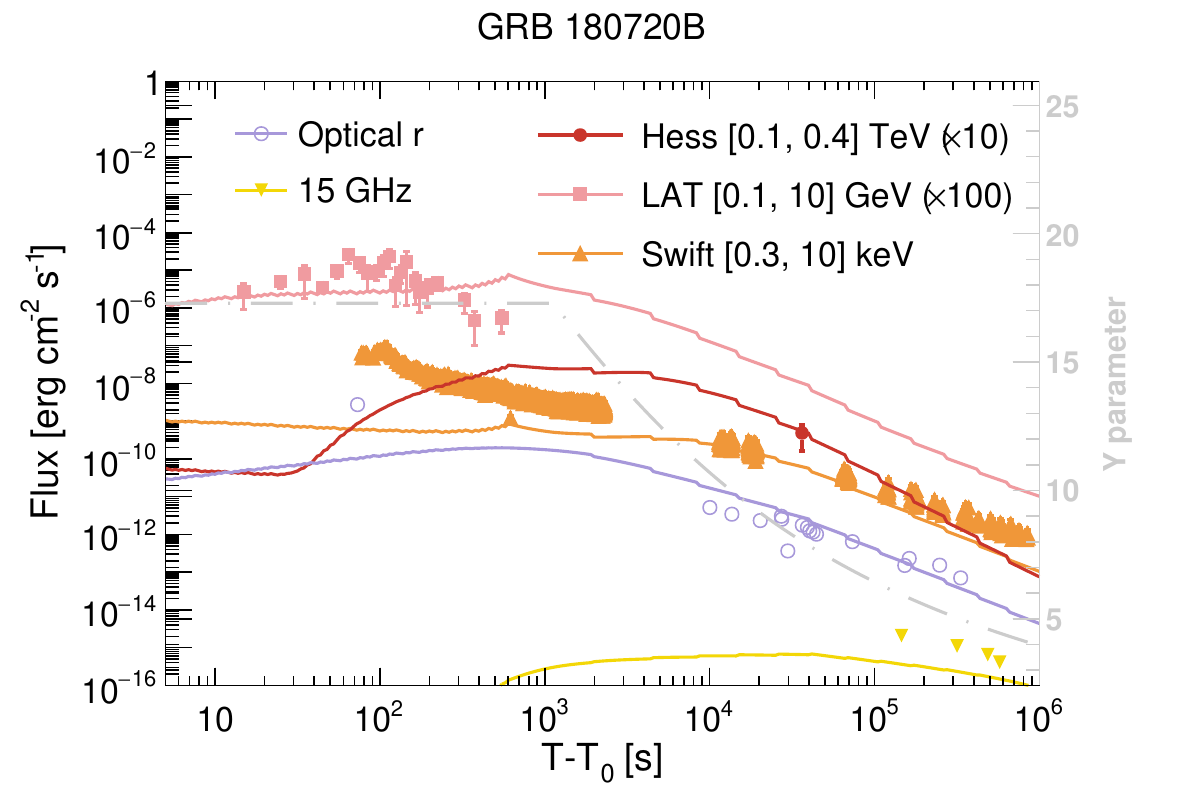}
\includegraphics[width=0.42\linewidth]{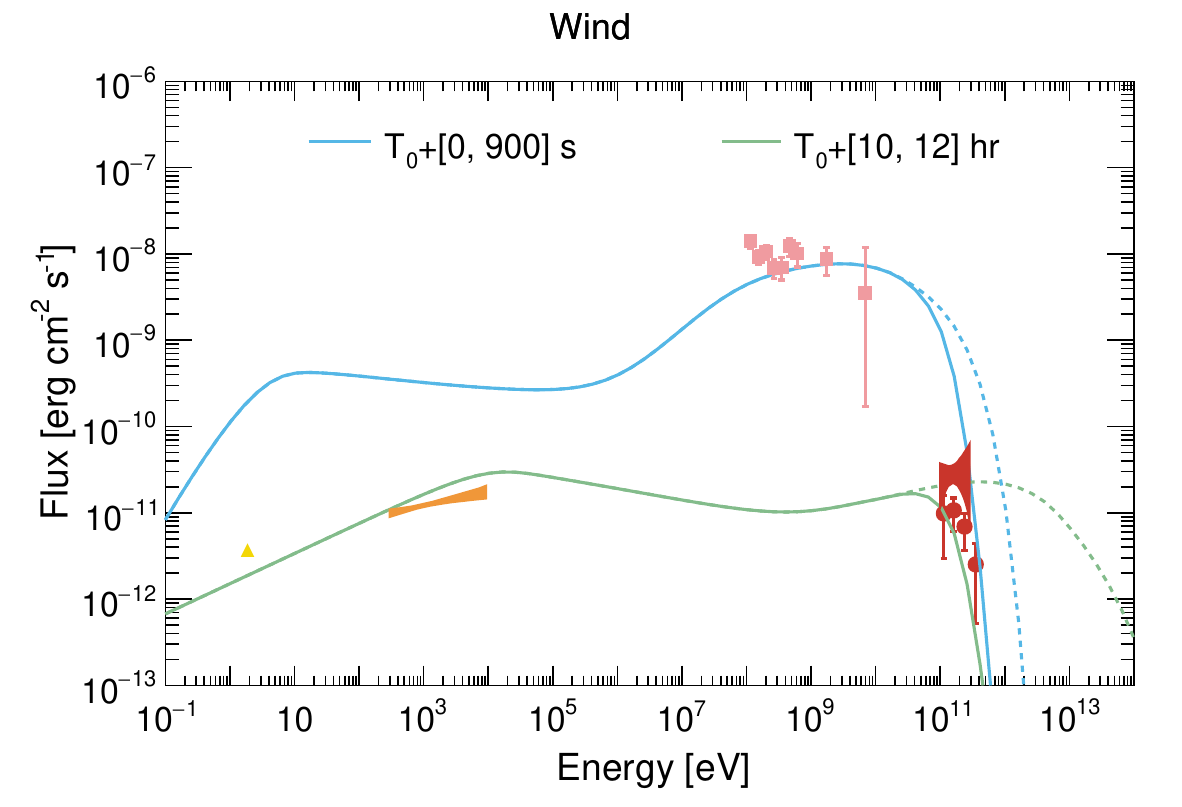}
\includegraphics[width=0.42\linewidth]{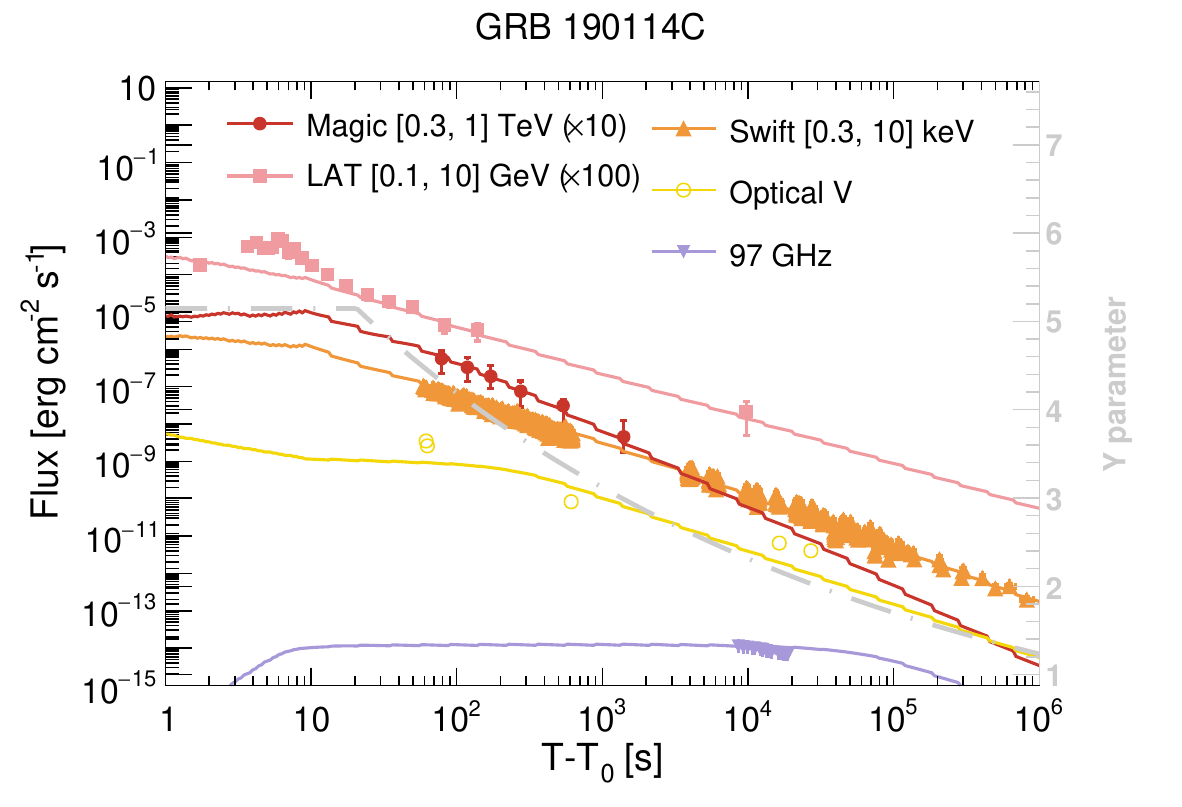}
\includegraphics[width=0.42\linewidth]{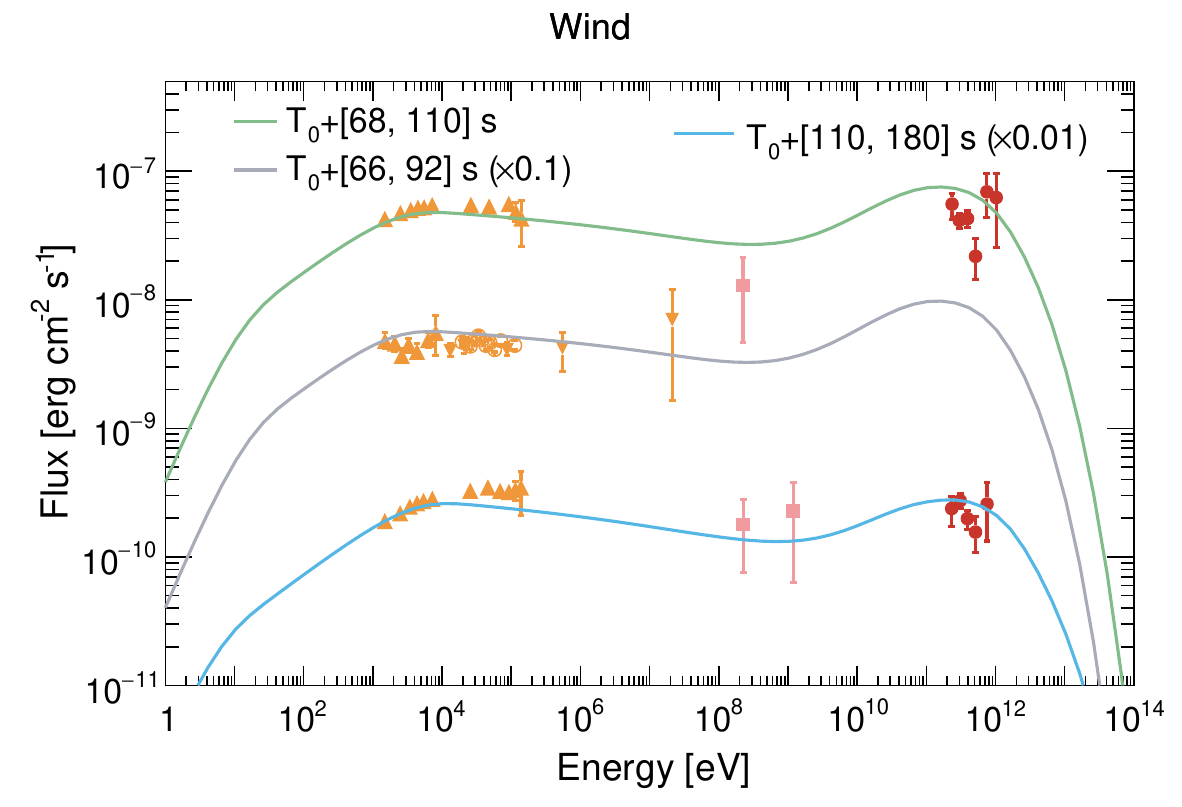}
\includegraphics[width=0.42\linewidth]{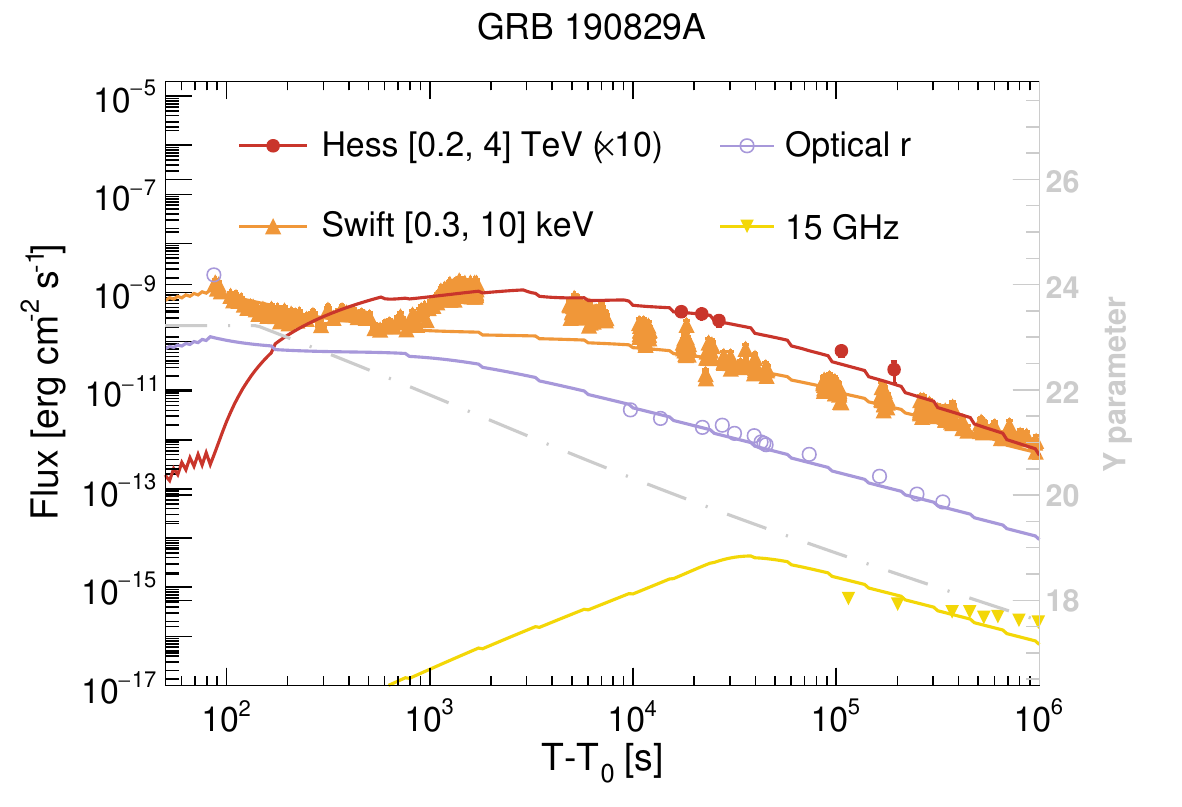}
\includegraphics[width=0.42\linewidth]{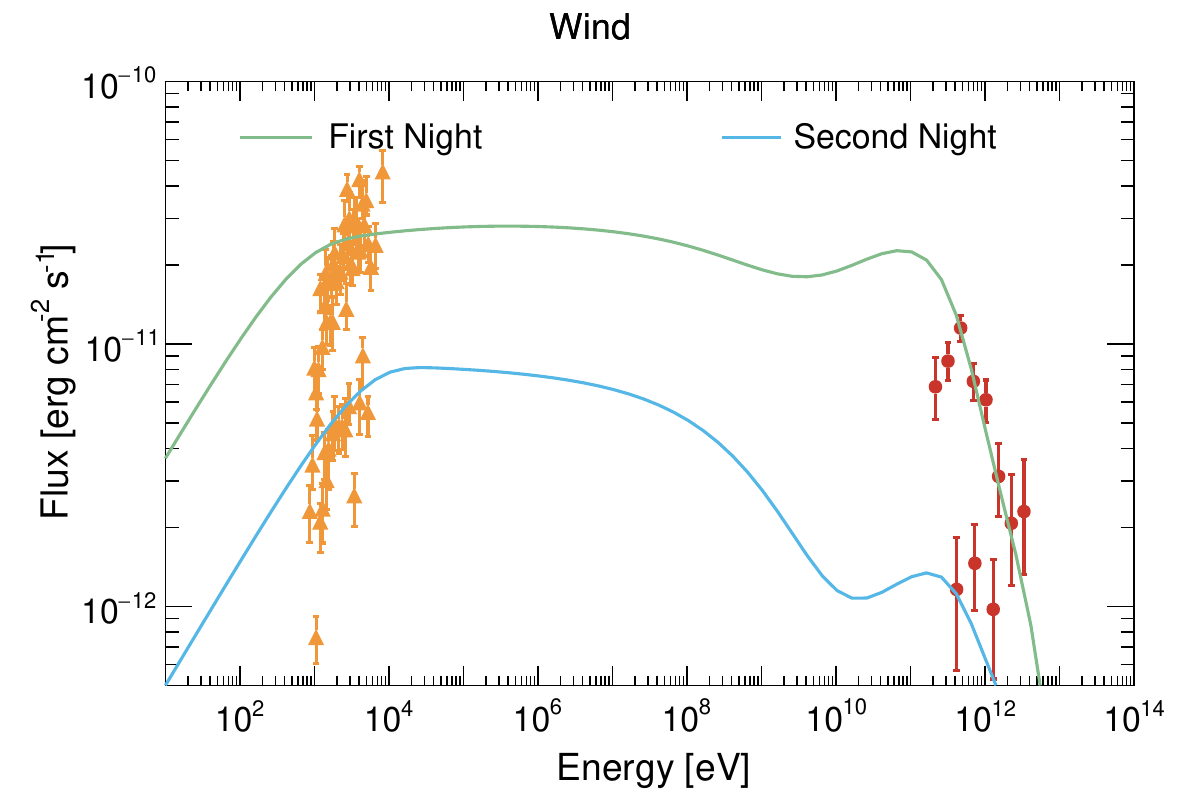}
\includegraphics[width=0.42\linewidth]{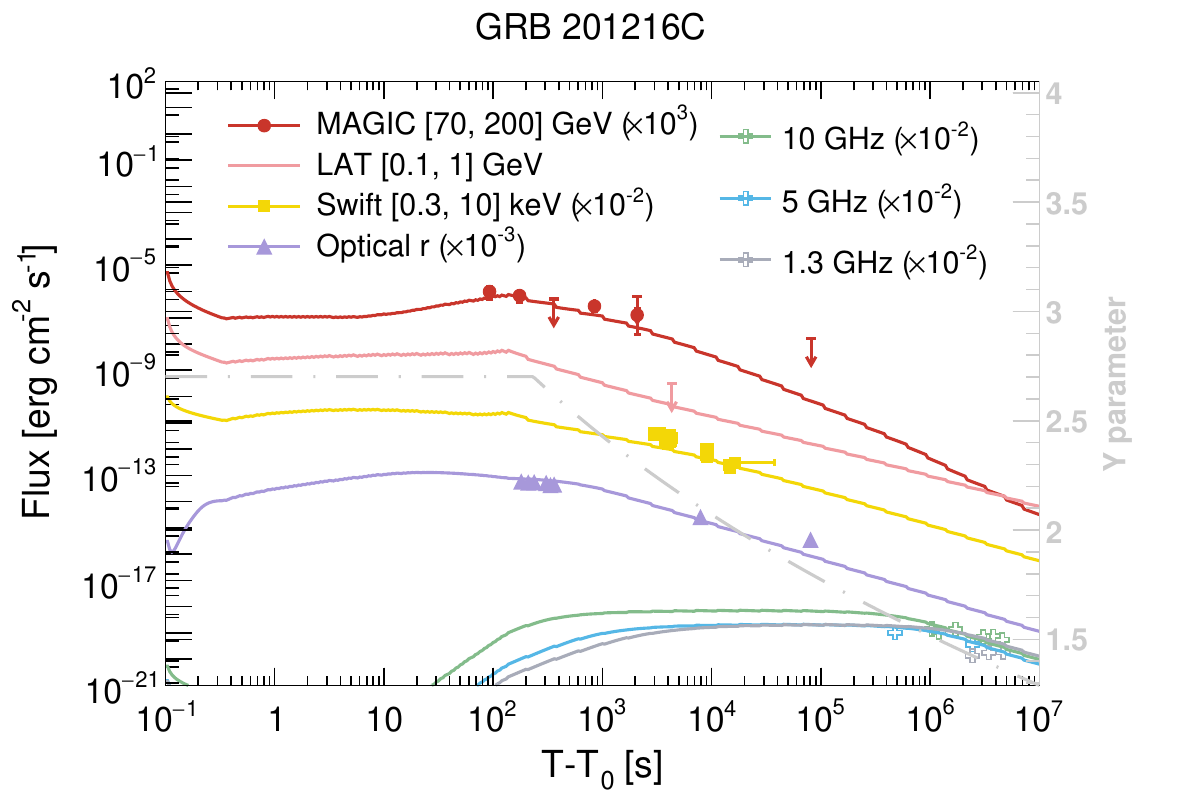}
\includegraphics[width=0.42\linewidth]{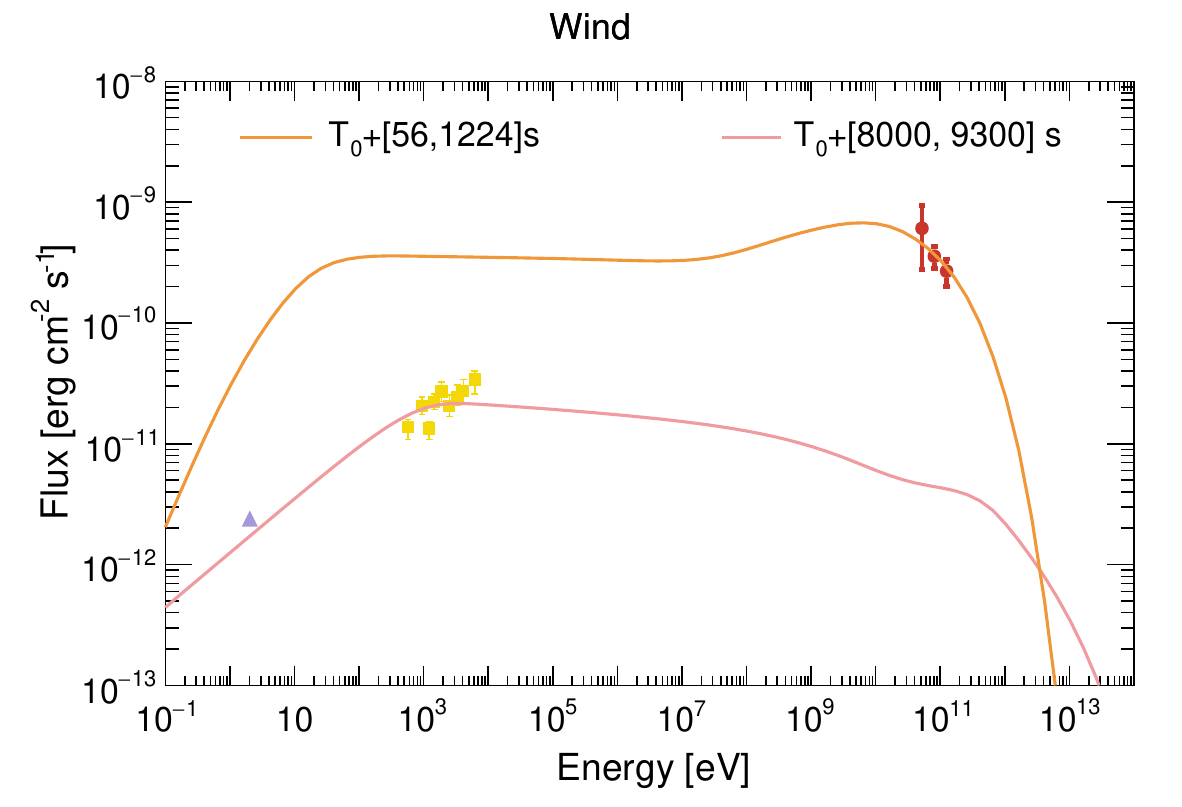}
\includegraphics[width=0.42\linewidth]{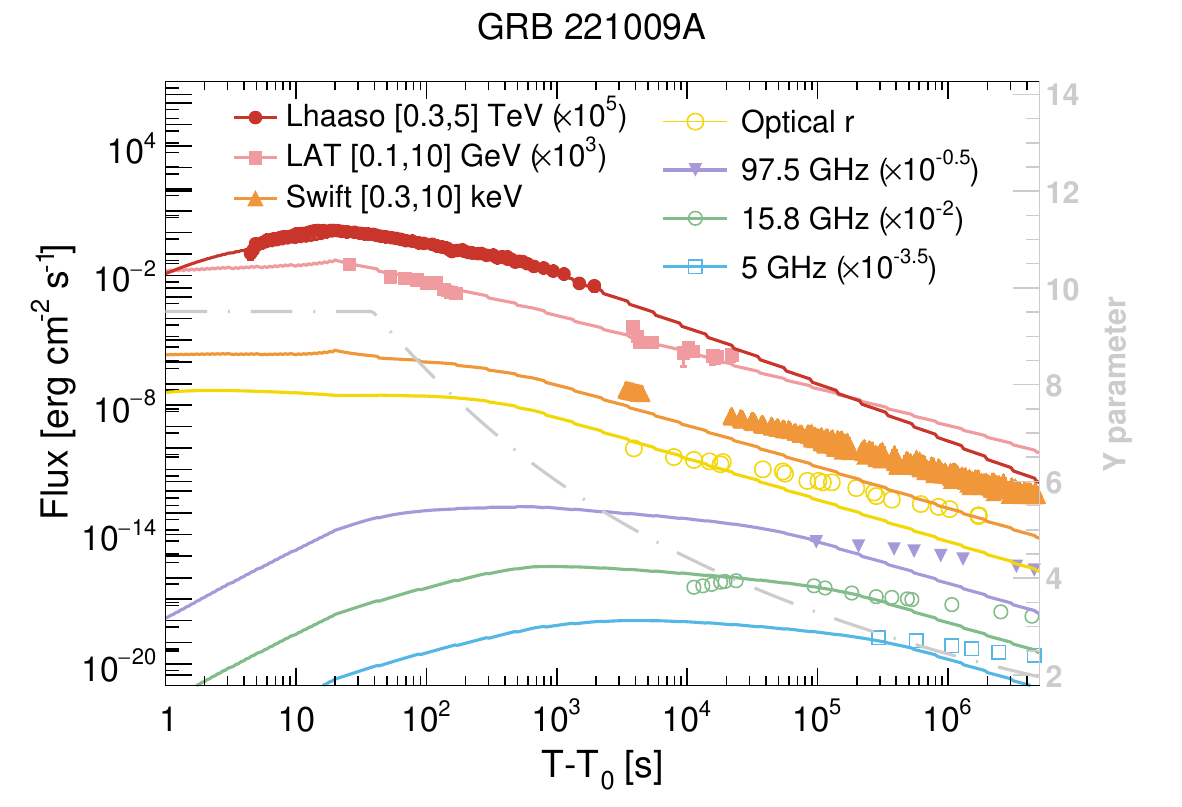}
\includegraphics[width=0.42\linewidth]{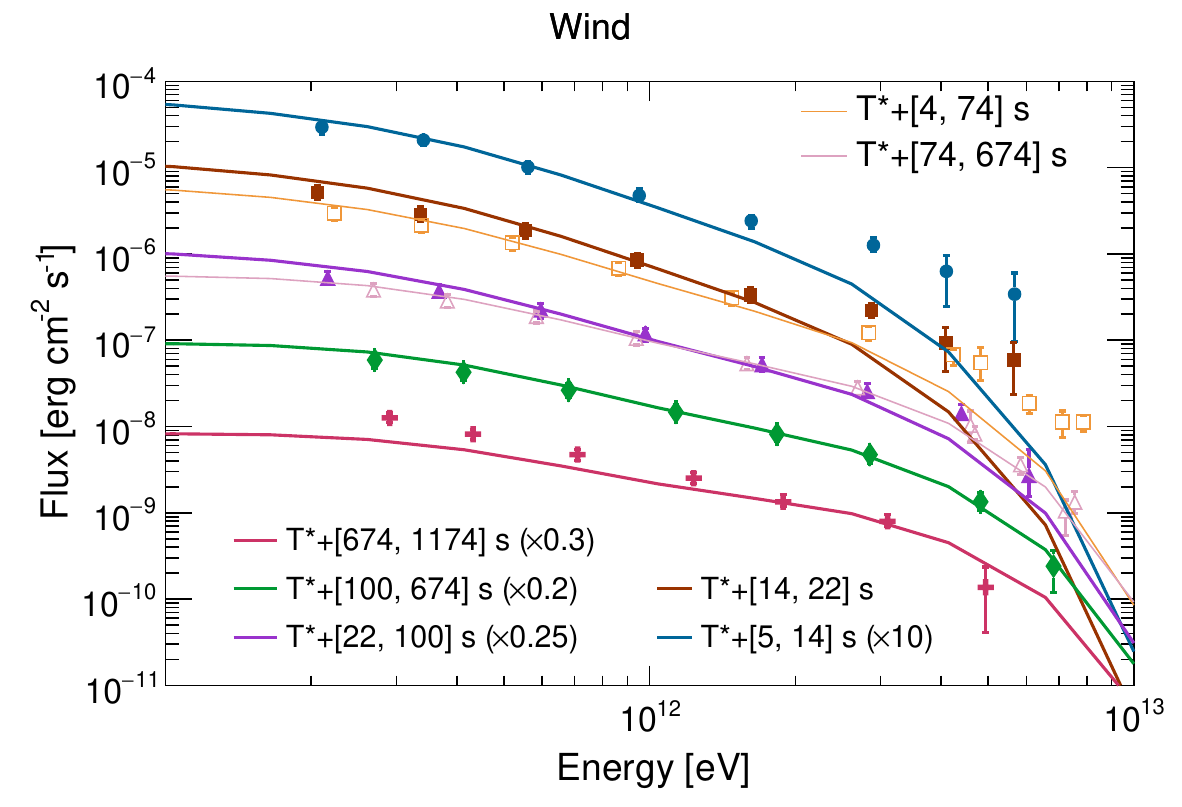}
\caption{{\scriptsize{Light curves (left) and SEDs (right) of SSC emission expected from a decelerating outflow in a stratified stellar wind medium. Lines in different colors are calculated with models, markers are observational data from different experiments. In right panels, the dashed lines in the highest energy range represents the flux after correcting EBL absorption. The observational data for GRB 180720B are obtained from \cite{2019Natur.575..464A,2019ApJ...885...29F,2019ApJ...885...29F,2024NatAs...8..134A}, for GRB 190114C are sourced from \cite{2019Natur.575..455M,2019ApJ...878L..26L,2019ApJ...883..162F}, for GRB 190829A are obtained from \cite{2021Sci...372.1081H,2023ApJ...947...84H}, for GRB 201216C are from \cite{2024MNRAS.527.5856A}, and for GRB 221009A are sourced from \cite{2023Sci...380.1390L,2023SciA....9J2778C,2024ApJ...962..115R}. The gray dashed lines in the left panel represent the time evolution of the Y parameter, their coordinates are the right y-axis of the graph.
}}}
\label{fig:wind}
\end{figure*}



\section{Results} 
In this section, we firstly present the modeling fitting results of five GRBs, under both ISM and wind-like cases. Subsequently, we compare the distribution of their model parameters with previous statistical analyses. Lastly, a potential correlation between the intrinsic energy budget in the keV-MeV and VHE energy range is explored.

\subsection{Model Fit to the Multiwavelength Data}

In a homogeneous ISM scenario, Figure \ref{fig:const} illustrates the modeling results of afterglow light curves (left) and SEDs (right), with detailed model parameters provided in Table \ref{tab:para}. The model calculates the time evolution of flux at different frequencies and compares it with observations spanning from radio to TeV range in the left panel. The model effectively explains the observed data points for these five GRBs, with the exception of the 97 GHz observation of GRB 190114C in the radio, where the model overpredicts the flux. Moreover, there are noticeable changes in the light curve slope for GRB 180270B and GRB 221009A, which could be attributed to jet breaks. Therefore, we include $ T_{break}=4\times10^{4}s$ for the former and $ T_{break}=670s$ for the latter \citep{2023Sci...380.1390L}.

The right panel of Figure \ref{fig:const} displays the spectra of GRBs, showing two distinct peaks. The low-energy peak is attributed to synchrotron emission, while the high-energy peak arises from SSC emission. The model fluxes are in good agreement with the energy spectra observed at various time intervals. It is worth noting that GRB 221009A is one of the most luminous GRBs, and its rapid flux increase post-onset led to saturation of nearly all space-based facilities. Here, we focus on presenting detailed observations in the TeV range.

The modeling and data comparison in the stratified wind-like environment are depicted in Figure \ref{fig:wind}. Contrasting the homogeneous scenario, the keV-range light curve of GRB 18072B under the wind-like conditions exhibits a slight underestimation compared to Swift observations before 10,000 seconds, suggesting the potential necessity for additional radiation mechanisms to reconcile the discrepancy. In the case of GRB 190114C, the wind-like model offers improved alignment with the observational data, successfully reproducing results even in the radio band. Notably, the highest energy segment of the energy spectra for GRB 221009A appears lower than the observations before 674 seconds; however, post this time interval, the model and observational data exhibit strong agreement.

In short, these five VHE-emission GRBs can be effectively described by the two common circumstellar media types, but subtle differences can still be observed when considering multi-band conformity. For GRBs 180720B and 221009A, the keV light curve of the former and the TeV energy spectrum of the latter suggest that a constant medium may provide a better fit to experimental data. In the case of GRB 190114C, the wind-like medium is more suitable than the constant medium, particularly excelling in alignment within the radio band. However, for GRB 190829A, although the models under both scenarios generally align with observational data, discrepancies exist in both the energy spectrum and keV light curve, especially evident in significant differences in the energy spectrum shape, indicating that alternative models may be required for a more comprehensive explanation.
\begin{table*}[!htp]
\caption{Model fit parameters}\label{tab:para}%
\begin{tabular*}{\linewidth}{@{\extracolsep\fill}cccccccc}
\hline
GRB  &                      & $log_{10}(E_k~[erg])$  & $log_{10}(\Gamma_{0})$  & $p$  & $log_{10}(\epsilon_B)$ & $ log_{10}(\epsilon_e)$ & $ log_{10}(n~[cm^{-3}])$ \\
\hline
180720B             &  ISM  & 53.75 & 2.4  &  2.1  & -3.5& -0.75 & -0.8 \\
                    &  Wind &    54                 &   1.6   &  2.3     &     -3.5                 &   -1  & 0.8\\
\hline
190114C             &  ISM  &  55 & 2.8  &  2.6 & -4 & -2 & -0.6 \\
                    &  Wind &      54               &    2.8  &    2.2   &         -3             &   -1.5  & 0 \\
\hline
190829A             & ISM   & 51.5 & 1.38  &  2.1  & -4  & -1 & 1.25 \\
                    &  Wind &        52             &    1.15  &     2.05  &         -4             &   -1.25  & 1 \\
\hline
201216C             & ISM   & 53.2 & 2.5  &  2.1  & -2.5  & -1 & -0.5 \\
                    & Wind &  53.2 & 2.  &   2.1  & -2    & -1  & 0 \\
\hline
221009A             & ISM &  55.3 & 2.9  &  2.12  & -3 & -1.65 & -1.64 \\
                    &  Wind &          55           &    2.35   &   2.35    &          -3.75            &   -1.75  & 0.6\\
\hline
\end{tabular*}
\end{table*}

\subsection{Comparative Analysis of Model Parameters}

\begin{figure*}[!htb]
\centering
\includegraphics[width=0.46\textwidth]{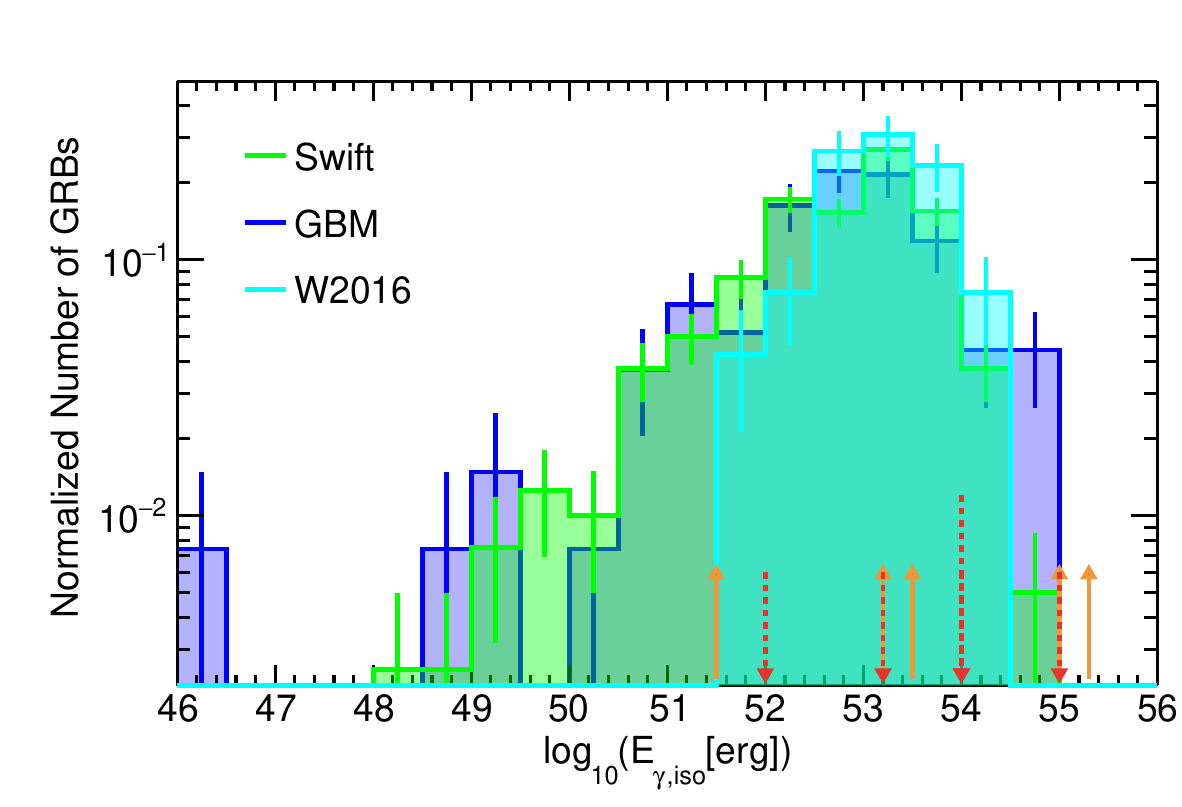}
\includegraphics[width=0.46\textwidth]{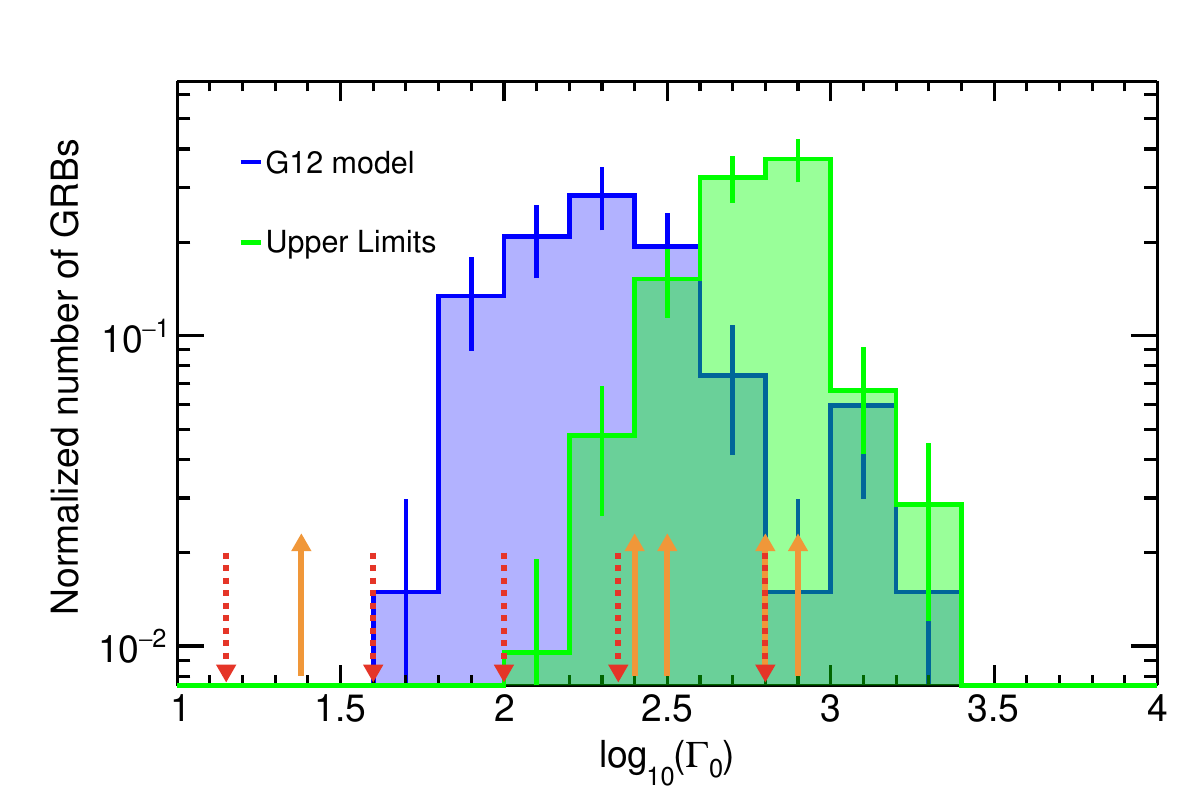}
\includegraphics[width=0.46\textwidth]{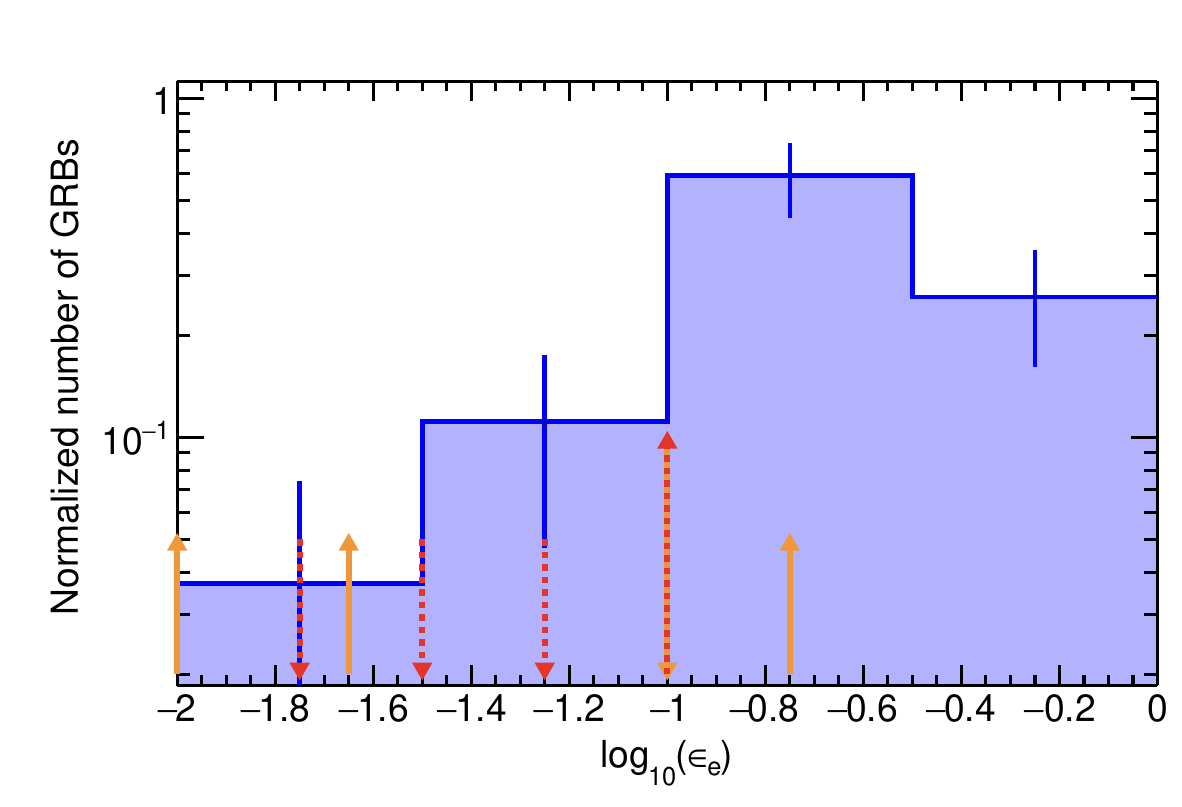}
\includegraphics[width=0.46\textwidth]{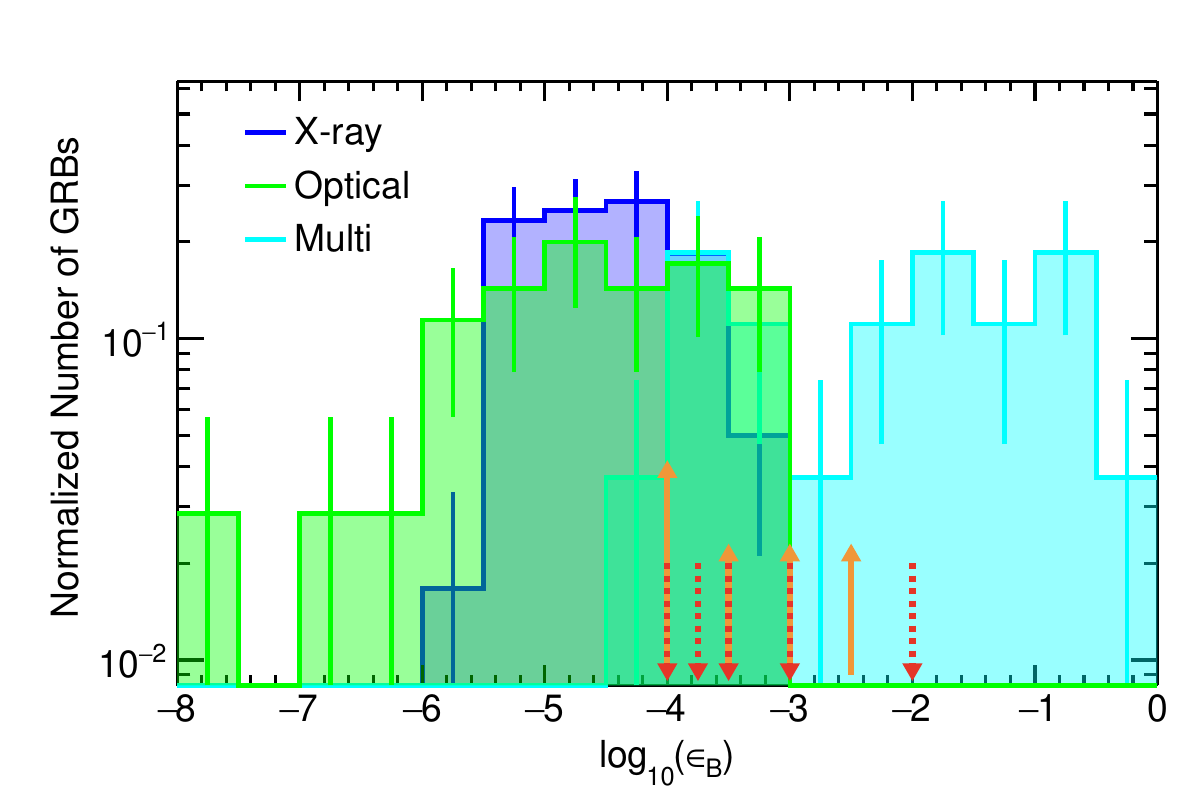}
\caption{Comparative analysis of key afterglow parameters of 5 VHE emission GRBs, with previous statistical GRB studies. The orange upward arrow represents the ISM scenario, while the red dashed line with downward arrows corresponds to the results in the case of a stellar wind. Top left: Distribution of total isotropic energy ($ E_{\gamma,iso}$) [adapted from \cite{2021ApJ...913...60P}]. Top right: Distribution of initial Lorentz factor ($ \Gamma_0$) [adapted from \cite{2018A&A...609A.112G}]. Bottom left: Distribution of electron energy fraction ($ \epsilon_e$) [adapted from \cite{2014ApJ...785...29S}]. Bottom right: Distribution of magnetic energy fraction ($ \epsilon_B$) [adapted from \cite{2014ApJ...785...29S}].}
\label{fig:para}
\end{figure*}
As previously mentioned, the afterglow model encompasses several crucial parameters that were statistically examined before VHE observations. With the inclusion of VHE radiation in this study, we now proceed to compare the features of the parameters derived in this research with those from prior studies, aiming to explore any disparities.

Figure \ref{fig:para} illustrates the model parameters obtained from the analysis of five GRBs, compared with the distribution of statistical studies on each parameter from the existing literature. The latter are shortly summarized as follows: 
\begin{enumerate}
    \item The Lorentz factor $\Gamma_0$ has been estimated using various methods, including the opacity method \citep{2009Sci...323.1688A,2014Sci...343...42A}, afterglow onset method \citep{2007A&A...469L..13M,2012MNRAS.420..483G,2018A&A...609A.112G}, and photosphere method \citep{2000ApJ...530..292M}. Among these methods, the afterglow onset method provides a more reliable estimation with fewer uncertainties. In this study, we rely on the distribution of Lorentz factors $\Gamma_0$ estimated from a robust sample of 66 GRBs with X-ray and optical observations, as well as data compiled from various previous studies \citep{2018A&A...609A.112G}.
    \item The electron's energy partition factor $\epsilon_e$ typically displays a remarkably narrow distribution \citep{2014ApJ...785...29S}, covering just one order of magnitude from 0.02 to 0.6. This distribution highlights that only a few GRBs have values of $\epsilon_e$ below 0.1.
    \item The distribution of the magnetic fields' energy partition factor $\epsilon_B$ proves to be considerably wider than that of $\epsilon_e$, lacking a clear range \citep{2014ApJ...785...29S}. Although previous studies on afterglow modeling have primarily focused on the range of $\epsilon_B$ between $\rm 10^{-4.5}$ and $\rm 10^0$, \cite{2014ApJ...785...29S} have derived their own results from X-ray and optical observations, which indicate that the distribution of $\epsilon_B$ is concentrated within the narrower range of $\rm 10^{-8}$ to $\rm 10^{-3}$.
\end{enumerate}

Figure \ref{fig:para} illustrates the profiles of four fundamental model parameters, namely $E_{\gamma,iso}$, $\Gamma_0$, $\epsilon_e$, and $\epsilon_B$, for five distinct GRBs, displaying their positions within the corresponding parameter distributions. The solid upper arrow represents the ISM scenario, while the dashed arrows depict the wind-like scenario. It is evident that there is no specific clustering in the first three parameters, despite the relatively low $\Gamma_0$ of 190829A under both environmental scenarios. However, the concentrated distribution of $\epsilon_B$ around $10^{-4}-10^{-3}$ is notable. This indicates that the inclusion of VHE observations in the multiwavelength dataset has led to a narrowing of the $\epsilon_B$ distribution, which was previously broad.

\subsection{Exploring Energy Budgets Correlation}

\begin{figure}[!htb]
\centering
\includegraphics[width=0.55\linewidth]{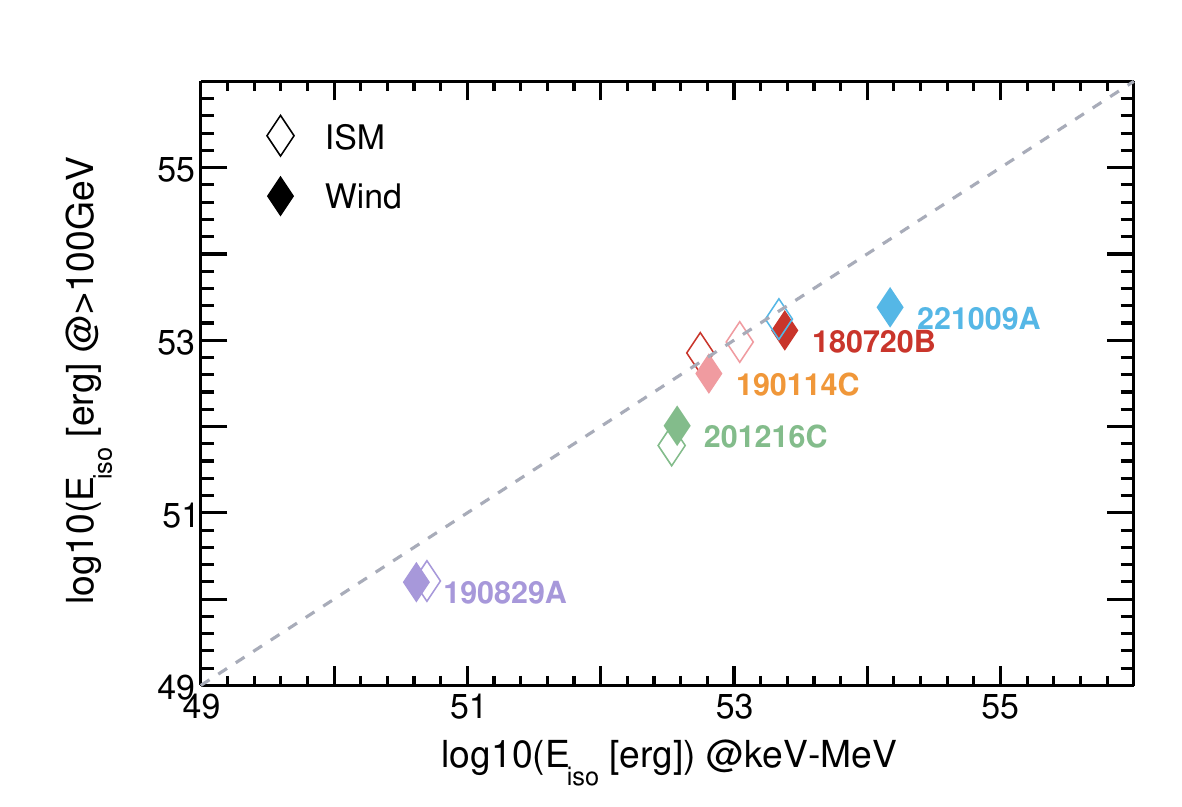}
\caption{Total energy budget ($E_{\gamma,iso}$,) distribution of 5 analyzed VHE GRBs in the keV-MeV and VHE energy ranges. The open (solid) diamonds represent results obtained from ISM (wind-like) environments. The dashed line represents a linear correlation.
}
\label{fig:Eiso}
\end{figure}


We further explore the potential relationship between the highest energy and lower energy ranges. Utilizing the expression $ E_{\gamma,iso}=4\pi D_L^2\int\int_{\nu_{min}}^{\nu_{max}} F(\nu, t) d\nu dt/(1+z)$, we have conducted calculations to determine the energy released within the keV-MeV and VHE ranges over a duration of $\rm10^{6}$ seconds. Figure \ref{fig:Eiso} illustrates the correlation between $ E_{keV,iso}$ and $ E_{VHE,iso}$ for the five observed GRBs, where the open (solid) diamonds represent results obtained from ISM (wind-like) environments. It is evident that the energy budget of keV-MeV and VHE emission in GRB 190114C under the wind-like scenario, and in GRB 190829A and GRB 221009A under the ISM case, exhibits nearly linear correlations, suggesting that their energy outputs are almost comparable. Conversely, the energy budget of other cases of this five GRBs, where the light curve or SED does not align with observations, shows deviation from the linear trend.

\begin{figure*}[!htp]
\centering
\includegraphics[width=0.49\linewidth]{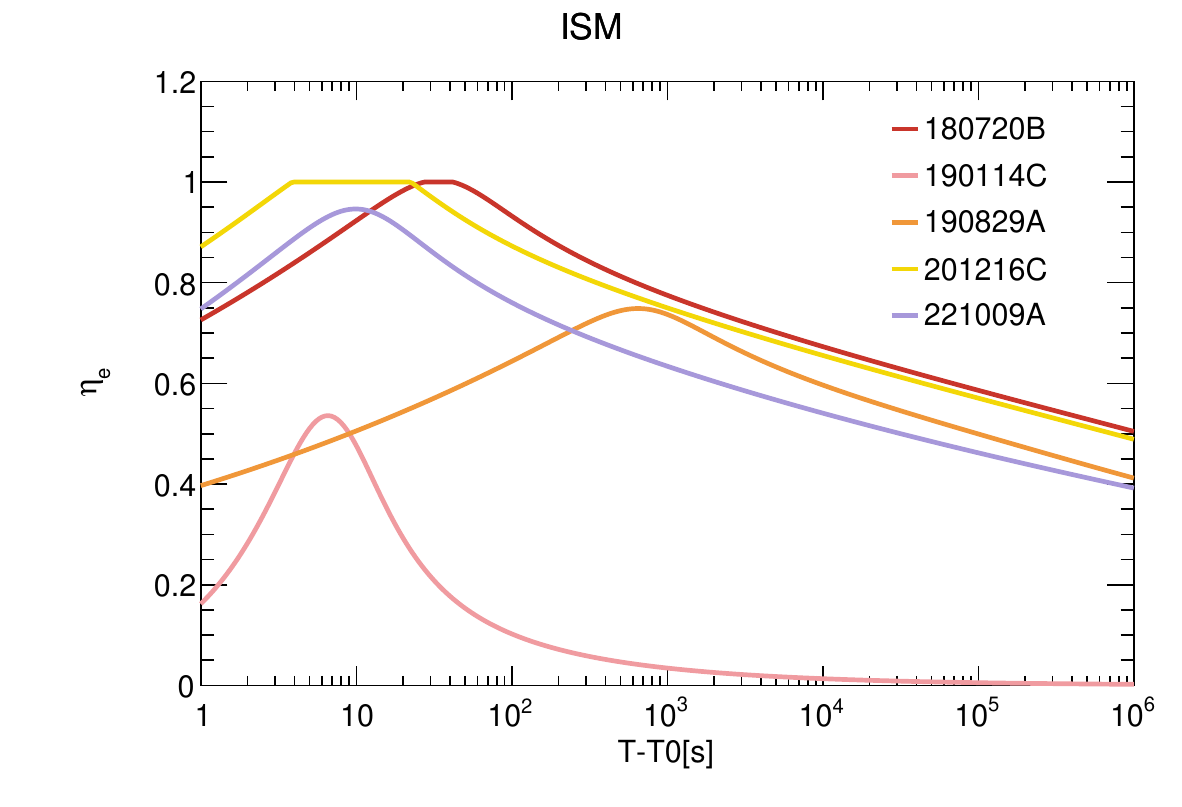}
\includegraphics[width=0.49\linewidth]{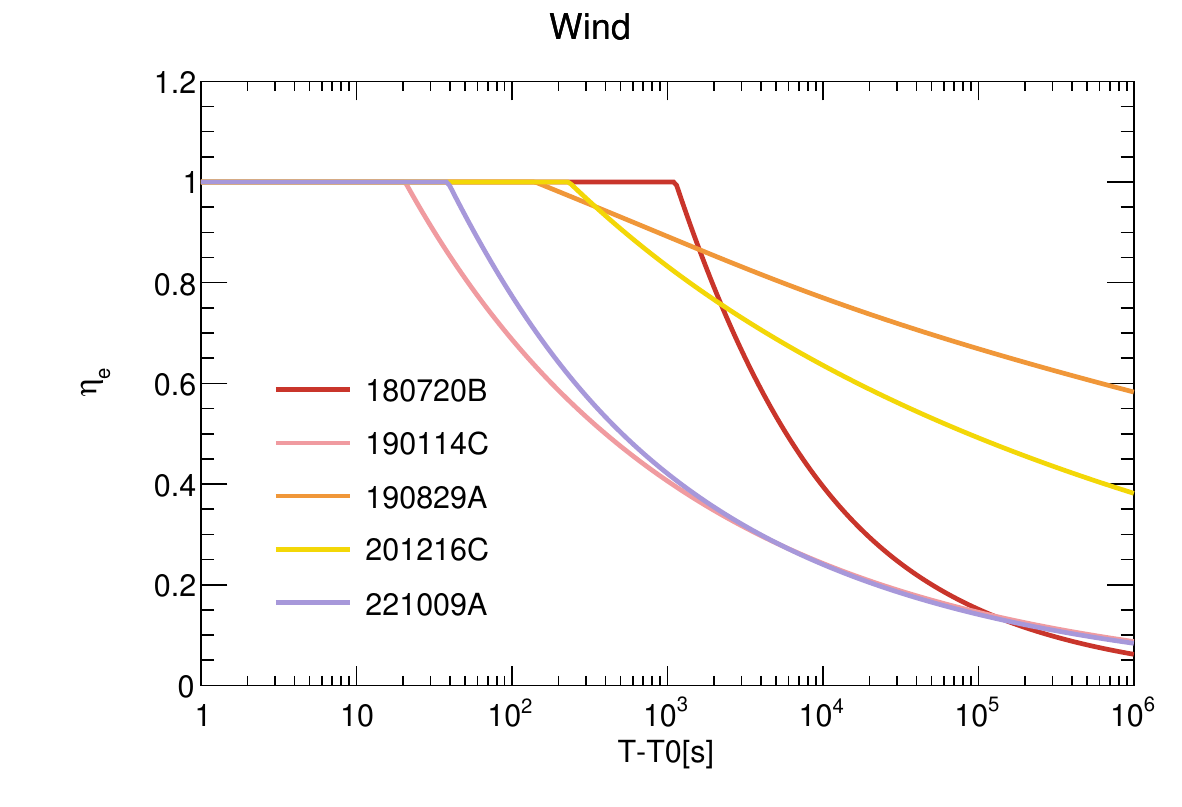}
\includegraphics[width=0.49\linewidth]{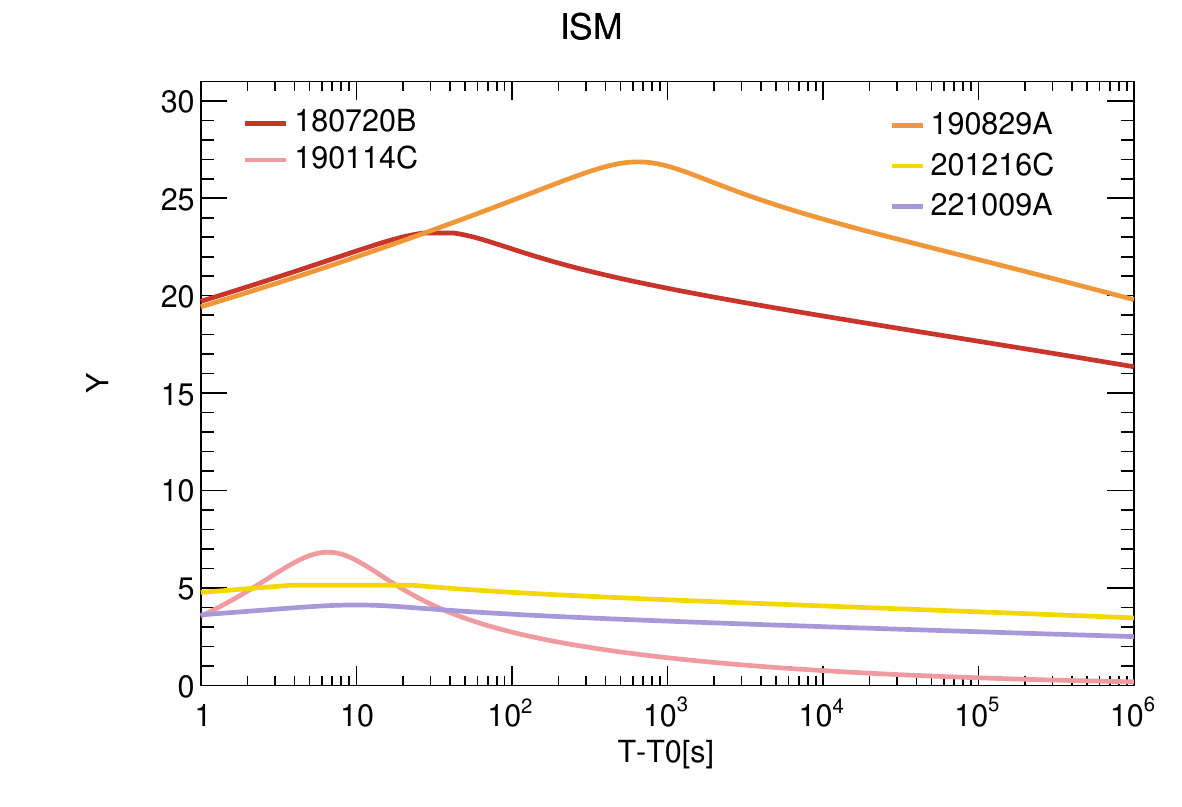}
\includegraphics[width=0.49\linewidth]{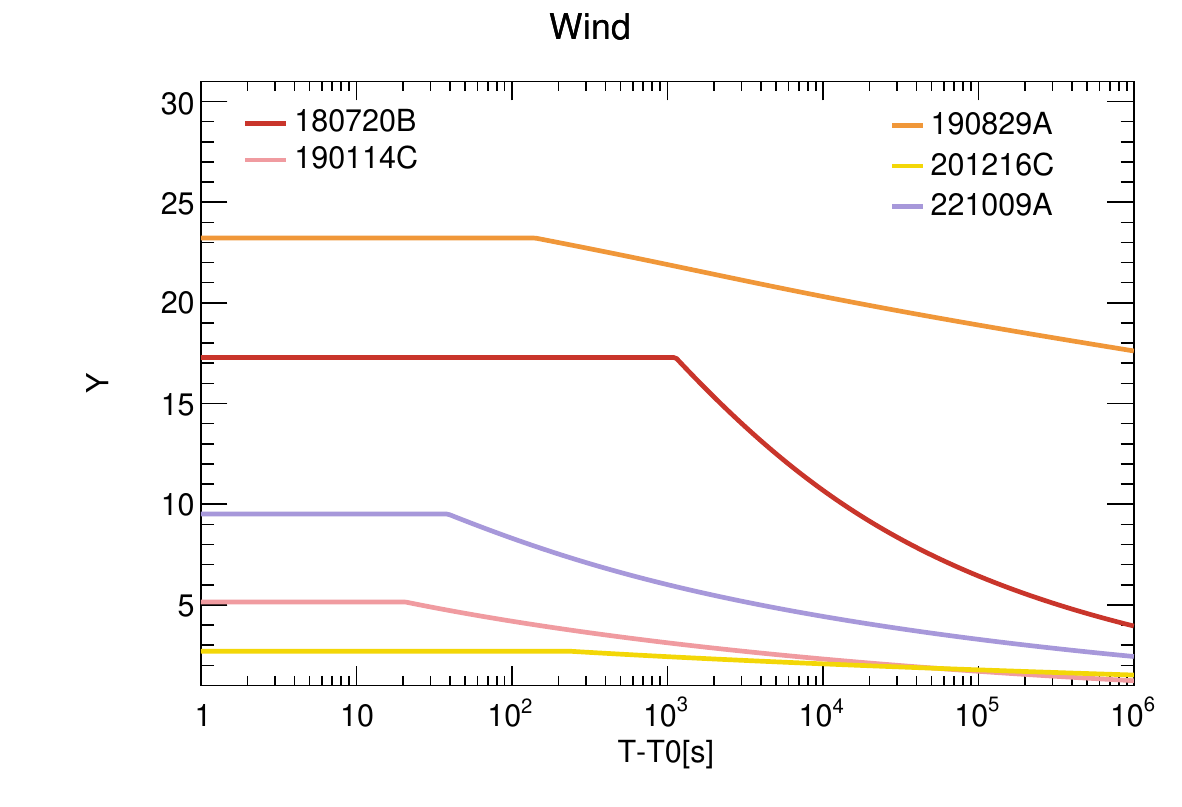}
\caption{ Time evolution of $\eta$ and Y parameters.
}
\label{fig:Y}
\end{figure*}

To make out the happenstance of VHE emission is induced by whether an selection of detection or connecting to the burst physical environment, GRB 180720B is taken as an example discussed in the following text. The afterglow peak time predicted by the external shock model is about $t_p \sim 30$ s, at a deceleration radius $R_{dec} \sim 10^{17}$ cm. For a moment after peaking time $t_p$, the standard slow cooling synchrotron (or SSC) takes the break power law form as
\begin{equation}
        f_{\nu} = f_{\nu,max}\left\{
\begin{aligned}
(\frac{\nu}{\nu_b})^{-\frac{p-1}{2}} & , & \nu_1 \le \nu \le \nu_b, \\
(\frac{\nu}{\nu_b})^{-\frac{p}{2}} & , & \nu_b \le \nu \le \nu_2.
\end{aligned}
\right.
\end{equation}
The differential energy emission on a explicit energy band $\left [ E_1,E_2 \right]$ (corresponding to $\left [ \nu_1,\nu_2 \right]$) is
\begin{equation}
\begin{split}
    \frac{dE}{dt}
    & = \int^{\nu_2}_{\nu_1}f_{\nu}d\nu \\
    & = f_{\nu,max}\nu_c\left\{\frac{2}{3-p}\left[1-\left(\frac{\nu_1}{\nu_b}\right)^{\frac{3-p}{2}}\right]+\frac{2}{2-p}\left[\left(\frac{\nu_2}{\nu_b}\right)^{\frac{2-p}{2}}-1\right]\right\}.
\end{split}
\end{equation}
The term in braces is the order of $\sim 1$, so the differential energy emission is $dE/dt\propto f_{\nu,max}\nu_b$, where $\nu_b$ is the peak energy of spectrum $\nu f_{\nu}$ and equals to cooling characteristic frequency $\nu_{c}^{syn}$ and
$\nu_{c}^{IC}$ for synchrotron and SSC emission respectively. Therefore, the differential energy ratio of SSC emission to synchrotron is
\begin{equation}
    \begin{split}
        \frac{dE_{SSC}}{dE_{syn}}
        & \approx \frac{f_{\nu,max}^{IC}\nu_{b}^{IC}}{f_{\nu,max}^{syn}\nu_{b}^{syn}} \\
        & \approx 4\sigma_Tn_eR\left(\frac{\gamma_c}{\gamma_m}\right)^{2-2p}\gamma_c^2.
    \end{split}
\end{equation}
Combine this equation with equation \ref{eq_gamma_m} and \ref{eq_gamma_c}, we can see that this ratio is related to every parameter of the external shock model
\begin{equation}
\begin{split}
\frac{dE_{SSC}}{dE_{syn}} =& 7.7\times10^{-8}\times10^{3(8p-14)}\left(\frac{3}{32\sigma_Tc}\right)^{4-2p} \\
& \times \left(\frac{m_p}{m_e}\right)^{4p-6}\left(\frac{p-2}{p-1}\right)^{2p-2}E_{53}^{\frac{1}{3}}\Gamma_{0,3}^{8p-\frac{44}{3}} \\
& \times \epsilon_B^{2p-4}\epsilon_e^{2p-2}n^{2p-\frac{10}{3}}\left(\frac{t}{t_p}\right)^{\frac{3}{2}-p}.
\end{split}
\end{equation}
For all those GRBs (for GRB 201216C, the main emission phase), the radiation processes are in the slow cooling regime, which makes the break frequencies in the equation above are $\nu_b^{syn} = \nu_c$ and $\nu_b^{IC} = 4\gamma_c^2\nu_cx_0$.
With the parameters of GRB 180720B, the coefficient before the time term is $\sim 2.6$. To survey the influence of various $p$ value to the ratio coefficient before the time term, we substitute $p^{\prime}$ for $p$: $p^{\prime}=p-2$. Then we divide the coefficient into two parts, one of them is $\Omega(p_i;p^{\prime})$ and it is sensitive to the value of $p$, while the other is $\Phi(p_i)$ and insensitive to $p$. The coefficient becomes
\begin{equation}
   coefficient =\Omega(p_i;p^{\prime})\Phi(p_i),
\end{equation}
where $\Omega(p_i;i)$ is
\begin{equation}
    \Omega(p_i;p^{\prime})=\left(10^{11.7}\Gamma_{0,3}^{8}\epsilon_B^{2}\epsilon_e^{2}n^{2}\right)^{p^{\prime}}
\end{equation}
and $\Phi(p_i)$ is
\begin{equation}
    \Phi(p_i)=10^{5.4}\left(\frac{p^{\prime}}{p^{\prime}+1}\right)^{2p^{\prime}+2}E_{53}^{\frac{1}{3}}\Gamma_{0,3}^{\frac{4}{3}}\epsilon_e^{2}n^{\frac{2}{3
    }}.
\end{equation}
Given the value of $\epsilon_B \sim 10^{-4}$, $\epsilon_B \sim 10^{-1}$, both terms are about the order of 1. But for each GRB, the break energy is might outrange of the observation band, or transits to lower energy in the main emission phase. This estimation is still qualitative and need to depends on the actual situation.
To explain the evolution of this ratio, we use the power law form of the characteristic Lorentz factor.  The time evolution of the variables are following a self-similar pattern: $\gamma_c \sim 10^5\left(\frac{t}{t_p}\right)^{\frac{1}{8}}$, $\gamma_m \sim 10^4\left(\frac{t}{t_p}\right)^{-\frac{3}{8}}$ and $R \sim 10^{17}\left(\frac{t}{t_p}\right)^{\frac{1}{4}}$ cm, which makes the differential ratio is: $\frac{dE^{SSC}}{dt}/\frac{dE^{syn}}{dt} = 0.6\left(\frac{t}{t_p}\right)^{\frac{3}{2}-p}$. In a short period $\left[t_p,10t_p\right]$, nearly half of gamma-ray energy is emitted. We take the energy emission rate of synchrotron as a power law for simplicity, $dE^{syn} = f(t)dt \propto \left(\frac{t}{t_p}\right)^{\alpha}dt$ with index $\alpha > 0$. The integration of the energy ratio on the time interval is
\begin{equation}
    \frac{E^{SSC}}{E^{syn}}=0.6\frac{t_2^{\alpha+\frac{5}{2}-p}-t_1^{\alpha+\frac{5}{2}-p}}{t_2^{\alpha+1}-t_1^{\alpha+1}},
\end{equation}
where $t_1 = 1$ and $t_2 = 10$. Ignoring the lower order terms, the energy ratio is $\frac{E^{SSC}}{E^{syn}}\sim0.15$, according with the model prediction of the ratio $\frac{E^{keV-MeV}}{E^{VHE}}=0.25$. It should be noted that we consider the  cooling characteristic frequencies of synchrotron and SSC are respectively in the keV-MeV and VHE band, for a relatively short time interval. This approximation is not so effective at the beginning and late afterglow, as both of them are not satisfied for GRB 180720B. However, our estimation can be of reference, because the most portion of kinematic energy of blastwave is lost at the early afterglow (hundreds of seconds after deceleration). We ignore the discussion about the time evolution of the break energy towards the observation energy band and leave it in the further work.

Figure \ref{fig:Y} presents the evolution of radiation efficiency $\eta$ and Y parameter for those five GRBs, indicating that the inverse Compton cooling is dominated at all the main emission phases. The time evolution behaviors are in line with \citet{2021MNRAS.504..528J}. For the transition time of inverse Compton cooling to synchrotron cooling, the Y parameters is approaching one, and the transition time expresses as
\begin{equation}
    t \sim \left(10^{3}\times2^{\frac{1}{2-p}}\times\frac{p-2}{p-1}\epsilon_B^{-\frac{3-p}{p-2}}\epsilon_e^{\frac{p-1}{p-2}}n\Gamma_{0,3}^4\right)^{2} \text{ s}.
\end{equation}
If accepting a value of $p = 2.5$, then the transition time is simplified as
\begin{equation}
    t \sim 10^{10}\epsilon_B^{-1}\epsilon_e^{3}n\Gamma_{0,3}^4 \text{s}.
\end{equation}

We can see that the transition time is very sensitive to the value of energy portion in electron accelerating and magnetic field. For the  medium density $n = 1$ cm$^{-3}$, initial bulk Lorenz factor $\Gamma = 200$, $\epsilon_e = 10^{-1}$, and a leak magnetic field case $\epsilon_B = 10^{-4}$, the transition time is $t \sim 2\times10^6$ s. The leading cooling mechanism is inverse Compton scattering for the whole afterglow phase. But for a strong magnetic field case $\epsilon_B = 10^{-1}$, the transition time is $t \sim 2$ s, corresponding to the synchrotron dominating situation. All our VHE GRBs afterglow explanation are inverse Compton dominating in the slow cooling phase, according with the standard external shock model. A comprehensive fitting was made to optical and X-ray afterglow, most of which are Swift GRBs \citep{Zhang2024}. Their results of fitting show a bimodal distribution of electron energy equipartition factor $\epsilon_e$ and a similar distribution of magnetic energy equipartition factor $\epsilon_B$ to this work. This discrepancy might be resulted by the deficiency of VHE emission in their GRB samples. Assuming the VHE emission is related to the X-ray band emission with a factor of $\sim 1$, the VHE GRBs year detection rate is still lower as $\sim$ 1 $\rm yr^{-1}$ \citep{Ashkar2024}. Only those nearly GRBs with less EBL absorption are capable to be detected. The VHE observations are also related to the ideal observation condition, which require low zenith angles, high duty cycle, short observational time delay, appropriate triggering time and so on. Adding all of those observation constraints can reduce the detection rate to a relative lower level.

\section{Summary}

The investigation of GRBs has experienced notable advancement in recent years, particularly within the keV-MeV energy spectrum. Multi-wavelength observations have emerged as a pivotal approach in elucidating the underlying physics of GRBs. The advent of VHE observations has offered valuable insights into the particle radiation mechanisms at play. In this study, we applied a standard one-zone synchrotron + SSC model to analyze five VHE GRBs. Notably, this model effectively reproduced the light curves and spectral energy distributions of the observed data. Specifically, we advocate for the SSC process in the case of GRB 190114C within a wind-like medium, while the VHE emissions observed in GRB 180720B and 221009A are attributed to interactions with the ISM. GRB 190829A, on the other hand, necessitates further investigation into additional radiation mechanisms for a comprehensive explanation. Despite the limited sample size, the microphysical parameters of VHE GRBs were found to be tightly constrained within a narrow range, particularly the $\epsilon_B$ parameter. This suggests that the inclusion of VHE observations significantly refines the parameter space and exerts a profound influence on associated physical parameters, such as the magnetic field strength. Furthermore, we have uncovered a robust and direct correlation between the keV-MeV and VHE energy bands, which might serve as a tool to differentiate radiation mechanisms. With the anticipation of detecting more VHE bursts, we look forward to further validating this correlation and enhancing our comprehension of GRB physics.

\acknowledgments
This work is supported by the National Natural Science Foundation of China (Nos. 12275279, 12405124), and the China Postdoctoral Science Foundation (No. 2023M730423).

\bibliographystyle{JHEP}
\bibliography{biblio.bib}
\end{document}